\documentclass[twocolumn]{aastex701} 
\usepackage{amsmath}
\usepackage{xcolor}
\usepackage{multirow}
\usepackage[T1]{fontenc}
\usepackage{booktabs}

\begin{document}

\title{Energy-Resolved Limits on Orbital X-ray Polarization Modulation in Cygnus X-1}

\author[orcid=0009-0002-2488-5272, gname=Sohee, sname='Chun']{Sohee Chun}
\affiliation{Department of Physics, McDonnell Center for the Space Sciences and Center for Quantum Sensors, Washington University in St. Louis, 1 Brookings Dr, Saint Louis, MO 63130, USA}
\email[show]{sohee.chun@wustl.edu}

\author[orcid=0000-0002-5488-1961, gname=Bert, sname='Vander Meulen']{Bert Vander Meulen} 
\affiliation{European Space Agency (ESA), European Space Research and Technology Centre (ESTEC), Keplerlaan 1, 2201 AZ Noordwijk, The Netherlands}
\email{bert.vandermeulen@esa.int}

\author[orcid=0000-0002-9705-7948, gname=Kun, sname='Hu']{Kun Hu}
\affiliation{Department of Physics, McDonnell Center for the Space Sciences and Center for Quantum Sensors, Washington University in St. Louis, 1 Brookings Dr, Saint Louis, MO 63130, USA}
\email{hkun@wustl.edu}

\author[orcid=0000-0002-1084-6507, gname=Henric, sname='Krawczynski']{Henric Krawczynski} 
\affiliation{Department of Physics, McDonnell Center for the Space Sciences and Center for Quantum Sensors, Washington University in St. Louis, 1 Brookings Dr, Saint Louis, MO 63130, USA}
\email{krawcz@wustl.edu}

\begin{abstract}
Reflection off the companion star and its focused stellar wind is predicted to modulate the X-ray polarization of black hole X-ray binaries at half the orbital period ($P_{\rm orb}/2$), with an energy-dependent amplitude.
We test this prediction against all publicly available \textit{IXPE} observations of Cygnus X-1, comprising 26 time bins of up to one day from 12 observation IDs spanning 2022--2024.
Since the normalized Stokes parameters correlate linearly with the spectral hardness ratio in all three energy bands (2--4, 4--6, and 6--8\,keV), we employ a simultaneous harmonic regression that decouples spectral variability from orbital modulation at both $P_{\rm orb}/2$ and $P_{\rm orb}$, complemented by direct fitting of 3D Monte Carlo radiative transfer stellar companion and wind-scattering templates.
After removing the spectral hardness trend, neither approach reveals statistically significant orbital modulation: permutation tests yield $p > 0.01$ in all bands, with 99\% confidence upper limits of 0.25\%, 0.63\%, and 1.96\% on the $P_{\rm orb}$ amplitude and 0.42\%, 0.97\%, and 2.78\% on the $P_{\rm orb}/2$ amplitude in the 2--4\,keV, 4--6\,keV, and 6--8\,keV bands, respectively.
The best-fit stellar companion and wind-scattering amplitude scaling factors in the three bands of $A = 0.67 \pm 0.85$, $0.97 \pm 0.62$, and $-1.06 \pm 1.11$ are consistent with a null result. 
These non-detections are sensitivity-limited, as the predicted stellar companion and wind-scattering RMS amplitudes in the three bands of $\approx$0.10\%, $\approx$0.33\%, and $\approx$0.49\% lie below the statistical noise floor of $\sim$0.18\%, $\sim$0.38\%, and $\sim$1.01\%.
We quantify the additional exposure required to detect the predicted signal and constrain the wind physics.
\end{abstract}

\keywords{\uat{Spectropolarimetry}{1973} --- \uat{High mass x-ray binary stars}{733} --- \uat{Stellar winds}{1636} --- \uat{Black hole physics}{159} --- \uat{Stellar mass black holes}{1611} --- \uat{Accretion}{14} --- \uat{High Energy astrophysics}{739}}

\section{Introduction}
Accreting black holes convert gravitational potential energy into bright radiation.
Revealing the physical properties of the accretion flows has been a long-standing problem in high-energy astrophysics as we would like to probe the physics of black hole accretion, the role of magnetic fields, the relativistic plasma processes at play, and the effects of strong gravity \citep[e.g.,][]{Narayan2005, Done2011, Krawczynski2012, Reynolds2021, Bambi2024}.
X-ray polarimetry offers new geometric insights beyond those available from spectral and timing studies alone.
The polarization degree (PD) and angle (PA) constrain the optical depth and geometry of the scattering medium and trace the orientation of the scattering plane \citep[e.g.,][]{Schnittman2009, Schnittman2010, Dovciak2011, 2022ApJ...934....4K}.

The launch of the \textit{Imaging X-ray Polarimetry Explorer (IXPE)} \citep{Weisskopf2022} in 2021 has revitalized X-ray polarimetry since the \textit{OSO-8} satellite \citep{Weisskopf1978}, enabling sensitive polarization measurements to constrain theoretical models.
\textit{IXPE} observations of black hole X-ray binaries (BHXRBs) reveal PDs typically ranging 1--8\% in the 2--8\,keV band, oftentimes increasing with energy.
The PAs seem to be aligned with radio jet axes favoring extended, disk-aligned coronal geometries over compact lamp-post configurations \citep[e.g.,][]{Krawczynski2022CygX1, Dovciak2024, Zhang2022}.

Cygnus X-1 is an ideal laboratory for exploring these effects.
It is a persistent, bright system hosting a $\sim$21\,$M_\odot$ black hole \citep{MillerJones2021} and an O-type supergiant companion, orbiting each other every 5.6 days \citep{Brocksopp1999}.
The system exhibits well-defined spectral state transitions \citep{DoneGierlinski2007, Grinberg2013} and strong orbital modulation in X-ray absorption \citep{Church2000, Poutanen2008, Hanke2009}.
Since it is a high-mass X-ray binary, accretion originates from a dense, structured stellar wind \citep{Gies1986a, Gies1986b}, creating a complex environment where both the accretion geometry and the line-of-sight scattering material vary with orbital phase \citep{Hanke2009, Miskovicova2016}.
Together, its brightness, dense focused wind, and phase-dependent scattering environment make Cygnus X-1 well suited for probing both intrinsic accretion variability and extrinsic geometric effects.

Disentangling geometric orbital modulation from intrinsic spectral variability is, however, a practical challenge.
The polarization properties of Cygnus X-1 depend strongly on the spectral state, and stochastic variations in spectral hardness can obscure orbital signatures in the Stokes parameters.
Indeed, during the \textit{IXPE} monitoring campaigns, Cygnus X-1 exhibited significant spectral variability on timescales comparable to the orbital period \citep{Krawczynski2022CygX1, Steiner2024, Jana2024}, making a joint analysis of spectral and orbital effects essential for a reliable interpretation of the polarization signatures.

\citet{VanderMeulen2026} developed a 3D Monte Carlo radiative transfer model of wind and companion scattering in Cygnus X-1 using the \texttt{SKIRT} code \citep{CampsBaes2020, VanderMeulen2023, VanderMeulen2024}.
Their model predicts a double-peaked orbital polarization modulation at $P_{\rm orb}/2$, with \textbf{minimum-to-maximum} PD amplitudes of $0.25$, $0.81$, and $1.24$ percentage points, corresponding to RMS amplitudes of the mean-subtracted Stokes parameters ($\sqrt{\langle \delta q^2 + \delta u^2 \rangle}$) of $0.10\%$, $0.33\%$, and $0.49\%$, in the 2--4, 4--6, and 6--8\,keV, respectively.
The amplitude increases with energy as soft X-rays are preferentially scattered by the dense wind near the black hole, suppressing net polarization, while hard X-rays penetrate deeper into the focused wind and reflect off the focused wind and the companion star, producing a more anisotropic scattering geometry and a larger net polarization modulation \citep{VanderMeulen2026}. \citet{Kravtsov2025} reported tentative variability at the fundamental period $P_{\rm orb}$ in the hard state, integrated over the full 2–-8 keV band.
The authors do not find obvious evidence for the $P_{\rm orb}/2$ reflection signature.
The results from \citet{VanderMeulen2026}
motivate a dedicated orbital phase, hardness ratio, and energy band resolved analysis of the entire {\it IXPE} dataset.

In this paper, we present such an analysis using all publicly available \textit{IXPE} observations of Cygnus X-1, comprising 26 time bins of up to one day from 12 observation IDs spanning 2022--2024.
We employ a simultaneous regression framework that models spectral hardness and orbital phase jointly, investigating the presence of a $P_{\rm orb}$ or $P_{\rm orb}/2$ signal, and we directly compare against the \texttt{SKIRT} templates of \citet{VanderMeulen2026}.
We validate the framework through injection-recovery simulations and a non-parametric permutation test, and quantify the sensitivity of the current dataset relative to the predicted signal amplitudes.
This analysis provides the first comparison of \citet{VanderMeulen2026} stellar companion and  wind-scattering predictions with the accumulated \textit{IXPE} dataset, evaluating the sensitivity of current observations relative to the predicted signal amplitudes.

The remainder of this paper is organized as follows. Section~\ref{sec:2.Data} describes the observational data and reduction procedure.
Section~\ref{sec:3.Dependence} characterizes the strong correlation between polarization and spectral hardness.
Section~\ref{sec:4.HarmReg} presents a model-independent search for polarization modulation at $P_{\rm orb}$ and $P_{\rm orb}/2$ using the simultaneous regression framework and its validation.
Section~\ref{sec:5.BertModel} presents the comparison with the \texttt{SKIRT} stellar companion and wind-scattering templates.
Section~\ref{sec:6.Discussion} discusses the sensitivity limits, physical interpretation, and prospects for future observations.

\section{Observations and Data Reduction} \label{sec:2.Data}
\textit{IXPE} carries three identical telescope modules, each consisting of a mirror module assembly (MMA) for focusing X-rays and a gas pixel detector (GPD), or detector unit (DU), for measuring polarization. Each GPD provides information about the position, energy, time, and polarization by tracking the photoelectrons produced by incoming X-rays. The three DUs operate simultaneously in the 2--8\,keV band, with a spatial resolution of $\sim 6.4''$ and an energy resolution of 0.54\,keV at 2\,keV (FWHM). Operating the three modules together improves polarimetric sensitivity and enables cross-checking \citep{Weisskopf2022}.

\textit{IXPE} observed Cygnus X-1 multiple times between 2022 and 2024. All publicly available observations were retrieved from the High Energy Astrophysics Science Archive (HEASARC) archive and analyzed using \texttt{ixpeobssim} v31.1.1 with the instrument response functions (\texttt{obssim20220702:v013}, \texttt{obssim20230702:v013}, and \texttt{obssim20240701:v013} for observations obtained in 2022, 2023, and 2024, respectively) \citep{Baldini2022}. For ObsID~01002901 (May 2022), obtained before the automatic energy calibration was implemented, we used energy-scale-corrected Level-2 files as in \citet{Krawczynski2022CygX1}.

Source events were extracted with \texttt{xpselect} from an $80''$ circular region centered on Cygnus X-1 (RA $\approx$299.6$^\circ$ and Dec$\approx$35.2$^\circ$, J2000), with the background estimated from a $150''$--$310''$ annular region. We applied the background rejection algorithm \citep{DiMarco2023} to all observations. Before binning, we identified the common good-time intervals (GTIs) across all three DUs by taking the intersection of their individual event time ranges, retaining only epochs when all three DUs were operating simultaneously. Background cubes were scaled by the ratio of the source-to-background extraction areas and then subtracted from the source cubes. The effective exposure for each bin was defined as the mean \texttt{LIVETIME} across the three DUs, as recorded in the event file headers. Each observation was divided into $\sim$1-day bins, yielding a total of 26 bins from 12 observation IDs. Note that ObsID~03002599 was split into two segments (03002599a and 03002599b) due to a 63-day gap between the observing epochs. 

\subsection{Spectral Characterization}
\begin{figure}
    \centering
    \includegraphics[width=\columnwidth]{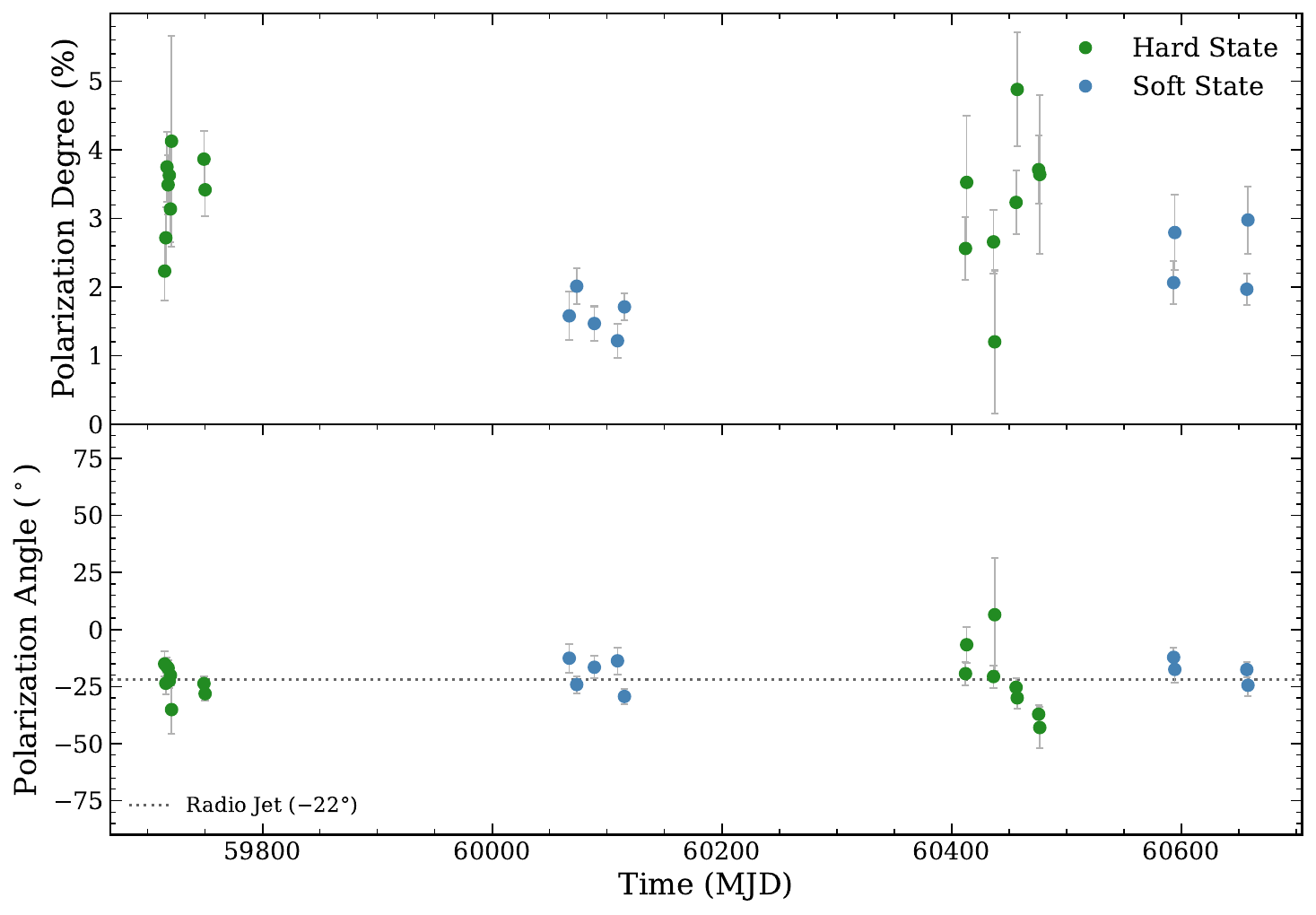}
    \caption{X-ray polarization in Cygnus X-1 from \textit{IXPE}. Top and bottom panels show the polarization degree (PD) and angle (PA) as a function of Modified Julian Date (MJD). Data points are color-coded by spectral state, defined by the hardness ratio (HR; hard state if $\mathrm{HR} \geq 0.46$ and soft state if $\mathrm{HR} < 0.46$). The PD is higher in the hard state than in the soft state, while the PA remains stable across states. All statistical calculations use the normalized Stokes parameters ($q$ and $u$) to avoid non-Gaussian errors and positive bias in PD \citep{Clarke1983}.}
    \label{fig:1_PDPAvsMJD}
\end{figure}

To characterize the spectral state of each bin, we computed the hardness ratio (HR), defined as
\begin{equation}
    \mathrm{HR} =  \frac{C(4\text{--}8\,\mathrm{keV})}
                        {C(2\text{--}4\,\mathrm{keV})},
\end{equation}
where $C(B)$ is the background-subtracted count rate in band $B$, extracted from Stokes $I$ cubes produced by \texttt{xpbin} in \texttt{PCUBE} mode, corrected for the effective area of the active detector units to account for epoch-dependent instrument response variations.
We define hard state if the HR is greater than 0.46 and soft state if less than 0.46.
While this boundary corresponds approximately to the canonical hard and soft state transition of Cygnus X-1, we note that HR as defined here is an \textit{IXPE}-specific quantity, dependent on the instrument's energy-dependent effective area even after correction, and this division is consistent with previous studies \citep[e.g.,][]{Kravtsov2025, Majumder2026}.
Note that since our analysis uses HR as a continuous covariate as described in Section~\ref{sec:4.HarmReg}, the choice of boundary only affects the state-averaged quantities of Table~\ref{tab:pd_pa_mean} and the state-resolved tests of Appendix~\ref{sec:appendixA}.
In any case the results are insensitive to the boundary because the measured HRs separate well with a gap between 0.391 and 0.520, so any boundary within this interval produces the identical partition and state-averaged values.
Across our sample, the HR spans from 0.145 to 0.704 (17 hard and 9 soft).
The evolution of these spectral states over the entire \textit{IXPE} campaign, along with their PD and PA, is illustrated in Figure~\ref{fig:1_PDPAvsMJD}.

\subsection{Polarization Measurements}
Polarization parameters were extracted in three energy bands (2--4, 4--6, and 6--8\,keV) using \texttt{xpbin} in \texttt{PCUBE} mode with the standard \textit{IXPE} event weighting \citep{Kislat2015}. Stokes $I$, $Q$, and $U$ maps were produced for each DU independently, background-subtracted after scaling by the ratio of source-to-background extraction areas, and finally combined across the three DUs using \texttt{xBinnedPolarization}\allowbreak\texttt{Cube.from\_file\_list} tool. Throughout this work, we report the normalized Stokes parameters:
\begin{equation}
    q \equiv \frac{Q}{I}, \qquad u \equiv \frac{U}{I},
\end{equation}
which have approximately Gaussian uncertainties, making them well-suited for the regression analysis described in Section~\ref{sec:4.HarmReg}.
The PD and PA are estimated via the standard relations\footnote{With $\mathrm{arctan2}$ defined as in \texttt{Python} or \texttt{C++}. Excel and Mathematica switch the arguments.} $\mathrm{PD} = \sqrt{q^2 + u^2}$ and $\mathrm{PA} = \frac{1}{2} \mathrm{arctan2}(u,q)$.

\subsection{Orbital Phase Assignment}
\begin{figure}
    \centering
    \includegraphics[width=\columnwidth]{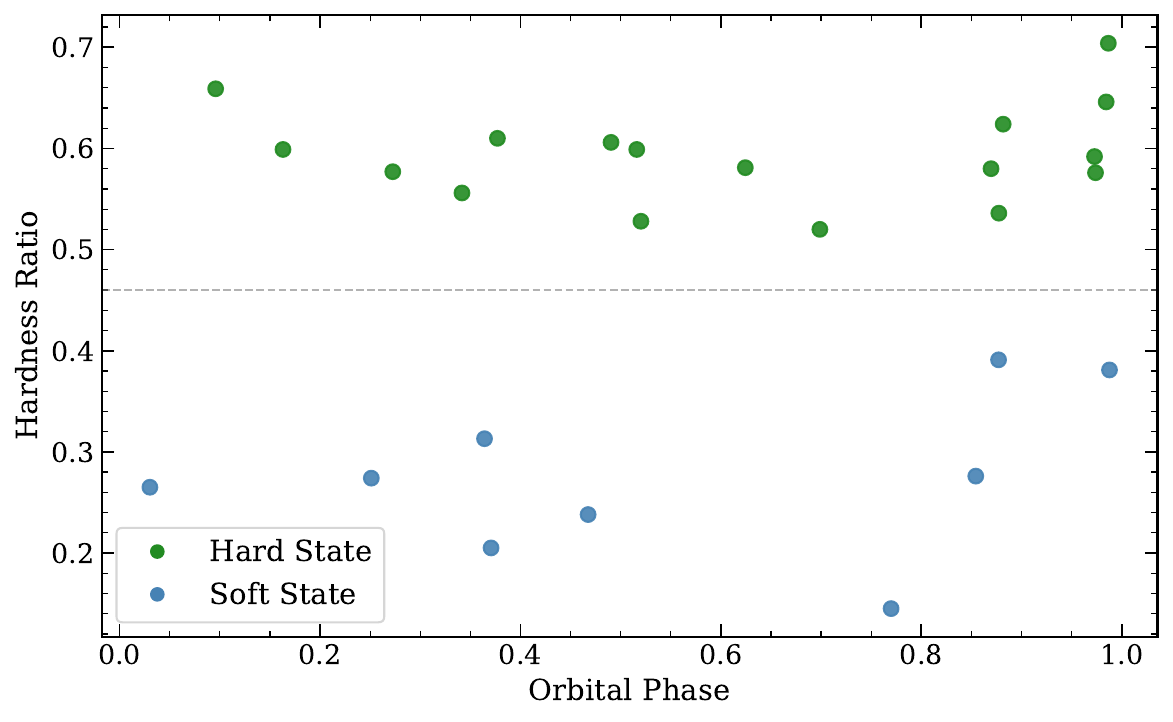}
    \caption{Hardness ratio (HR) as a function of orbital phase for the 26 bins. The absence of significant correlation shows that the spectral state sampling is approximately uniform across orbital phases, validating the simultaneous regression approach of Section~\ref{sec:4.HarmReg}.}
    \label{fig:2_HRvsOP}
\end{figure}

Orbital phases were computed using the ephemeris of $P_{\rm orb} = 5.599829$ days and $T_0 = \mathrm{JD}\,2441874.707$, with $\phi=0$ at superior conjunction of the black hole \citep{Brocksopp1999}.
Our 26 temporal bins span the full phase range $0.030 \leq \phi \leq 0.988$, providing reasonably uniform phase coverage, but the range $\phi \approx 0.5$--$0.8$ remains comparatively under-sampled (Figure~\ref{fig:2_HRvsOP}).
Given $T_0 = \mathrm{JD}\,2441874.707 \pm 0.009$ and $P_{\rm orb} = 5.599829 \pm 0.000016$\,days \citep{Brocksopp1999}, our 2022--2024 observations span $N_{\rm cycle} \approx 3186$ to $3354$ elapsed cycles since $T_0$.
The period uncertainty contributes $N_{\rm cycle} \times \sigma_P/P_{\rm orb} = 0.0091$--$0.0096$ and the epoch uncertainty $\sigma_{T_0}/P_{\rm orb} = 0.0016$.
Combined, these give $\sigma_\phi = 0.0092$--$0.0097$, or 1.24--1.31\,hr, corresponding to $\lesssim$5\% of our $\sim$1-day bin width and therefore negligible for this analysis.
The full dataset is presented in Table~\ref{tab:observations}.

\begin{table*}
    \centering
    \hspace{-2cm}
    \resizebox{\textwidth}{!}{
    \begin{tabular}{lccccccccc}
    \hline
    OBSID & Bin & $\phi$ & HR & \multicolumn{2}{c}{2--4 keV} & 
    \multicolumn{2}{c}{4--6 keV} & \multicolumn{2}{c}{6--8 keV} \\
     & & & & $q$ & $u$ & $q$ & $u$ & $q$ & $u$ \\
    \hline
    01002901 & 0 & 0.984 & 0.646 & $0.019 \pm 0.004$ & $-0.011 \pm 0.004$ & $0.030 \pm 0.007$ & $-0.038 \pm 0.007$ & $0.046 \pm 0.019$ & $-0.046 \pm 0.019$ \\
    ~ & 1 & 0.163 & 0.599 & $0.019 \pm 0.005$ & $-0.020 \pm 0.005$ & $0.035 \pm 0.008$ & $-0.025 \pm 0.008$ & $0.004 \pm 0.020$ & $-0.033 \pm 0.020$ \\
    ~ & 2 & 0.342 & 0.556 & $0.032 \pm 0.005$ & $-0.020 \pm 0.005$ & $0.024 \pm 0.010$ & $-0.028 \pm 0.010$ & $0.017 \pm 0.024$ & $-0.020 \pm 0.024$ \\
    ~ & 3 & 0.520 & 0.528 & $0.029 \pm 0.004$ & $-0.019 \pm 0.004$ & $0.024 \pm 0.008$ & $-0.039 \pm 0.008$ & $0.055 \pm 0.021$ & $-0.069 \pm 0.021$ \\
    ~ & 4 & 0.699 & 0.520 & $0.026 \pm 0.004$ & $-0.026 \pm 0.004$ & $0.035 \pm 0.008$ & $-0.022 \pm 0.008$ & $0.022 \pm 0.020$ & $-0.051 \pm 0.020$ \\
    ~ & 5 & 0.877 & 0.536 & $0.024 \pm 0.005$ & $-0.020 \pm 0.005$ & $0.041 \pm 0.009$ & $-0.037 \pm 0.009$ & $0.049 \pm 0.023$ & $-0.031 \pm 0.023$ \\
    ~ & 6 & 0.973 & 0.592 & $0.014 \pm 0.015$ & $-0.039 \pm 0.015$ & $-0.018 \pm 0.027$ & $-0.040 \pm 0.027$ & $0.003 \pm 0.070$ & $-0.108 \pm 0.070$ \\
    01250101 & 0 & 0.096 & 0.659 & $0.026 \pm 0.004$ & $-0.028 \pm 0.004$ & $0.032 \pm 0.007$ & $-0.034 \pm 0.007$ & $0.028 \pm 0.017$ & $-0.044 \pm 0.017$ \\
    ~ & 1 & 0.273 & 0.577 & $0.019 \pm 0.004$ & $-0.028 \pm 0.004$ & $0.021 \pm 0.007$ & $-0.021 \pm 0.007$ & $0.036 \pm 0.017$ & $-0.018 \pm 0.017$ \\
    02008201 & 0 & 0.854 & 0.276 & $0.014 \pm 0.004$ & $-0.007 \pm 0.004$ & $0.022 \pm 0.008$ & $-0.027 \pm 0.008$ & $0.065 \pm 0.023$ & $-0.010 \pm 0.023$ \\
    02008301 & 0 & 0.030 & 0.265 & $0.013 \pm 0.003$ & $-0.015 \pm 0.003$ & $0.031 \pm 0.006$ & $-0.017 \pm 0.006$ & $0.025 \pm 0.018$ & $-0.043 \pm 0.018$ \\
    02008401 & 0 & 0.770 & 0.145 & $0.012 \pm 0.003$ & $-0.008 \pm 0.003$ & $0.026 \pm 0.008$ & $-0.015 \pm 0.008$ & $0.049 \pm 0.025$ & $-0.064 \pm 0.025$ \\
    02008501 & 0 & 0.371 & 0.205 & $0.011 \pm 0.003$ & $-0.006 \pm 0.003$ & $0.013 \pm 0.007$ & $-0.027 \pm 0.007$ & $0.032 \pm 0.019$ & $0.006 \pm 0.019$ \\
    02008601 & 0 & 0.468 & 0.238 & $0.009 \pm 0.002$ & $-0.015 \pm 0.002$ & $0.016 \pm 0.005$ & $-0.025 \pm 0.005$ & $-0.002 \pm 0.014$ & $-0.025 \pm 0.014$ \\
    03002201 & 0 & 0.516 & 0.599 & $0.020 \pm 0.005$ & $-0.016 \pm 0.005$ & $0.021 \pm 0.008$ & $-0.028 \pm 0.008$ & $-0.003 \pm 0.020$ & $-0.080 \pm 0.020$ \\
    ~ & 1 & 0.624 & 0.581 & $0.034 \pm 0.010$ & $-0.008 \pm 0.010$ & $0.026 \pm 0.017$ & $-0.029 \pm 0.017$ & $0.041 \pm 0.044$ & $-0.015 \pm 0.044$ \\
    03003101 & 0 & 0.870 & 0.580 & $0.020 \pm 0.005$ & $-0.018 \pm 0.005$ & $0.028 \pm 0.008$ & $-0.015 \pm 0.008$ & $0.050 \pm 0.021$ & $-0.018 \pm 0.021$ \\
    ~ & 1 & 0.974 & 0.576 & $0.012 \pm 0.010$ & $0.003 \pm 0.010$ & $0.017 \pm 0.019$ & $-0.030 \pm 0.019$ & $0.053 \pm 0.048$ & $0.004 \pm 0.048$ \\
    03010001 & 0 & 0.377 & 0.610 & $0.021 \pm 0.005$ & $-0.025 \pm 0.005$ & $0.027 \pm 0.008$ & $-0.044 \pm 0.008$ & $0.057 \pm 0.020$ & $-0.010 \pm 0.020$ \\
    ~ & 1 & 0.490 & 0.606 & $0.025 \pm 0.008$ & $-0.042 \pm 0.008$ & $0.018 \pm 0.015$ & $-0.051 \pm 0.015$ & $-0.020 \pm 0.037$ & $-0.021 \pm 0.037$ \\
    03010101 & 0 & 0.882 & 0.624 & $0.010 \pm 0.005$ & $-0.036 \pm 0.005$ & $0.022 \pm 0.009$ & $-0.035 \pm 0.009$ & $0.051 \pm 0.022$ & $-0.063 \pm 0.022$ \\
    ~ & 1 & 0.987 & 0.704 & $0.003 \pm 0.012$ & $-0.036 \pm 0.012$ & $0.034 \pm 0.019$ & $-0.039 \pm 0.019$ & $0.072 \pm 0.048$ & $-0.026 \pm 0.048$ \\
    03002599a & 0 & 0.877 & 0.391 & $0.019 \pm 0.003$ & $-0.009 \pm 0.003$ & $0.027 \pm 0.007$ & $-0.040 \pm 0.007$ & $0.031 \pm 0.018$ & $-0.022 \pm 0.018$ \\
    ~ & 1 & 0.988 & 0.381 & $0.023 \pm 0.006$ & $-0.016 \pm 0.006$ & $0.029 \pm 0.011$ & $-0.033 \pm 0.011$ & $0.029 \pm 0.032$ & $-0.060 \pm 0.032$ \\
    03002599b & 0 & 0.251 & 0.274 & $0.016 \pm 0.002$ & $-0.011 \pm 0.002$ & $0.024 \pm 0.006$ & $-0.026 \pm 0.006$ & $0.034 \pm 0.016$ & $-0.026 \pm 0.016$ \\
    ~ & 1 & 0.364 & 0.313 & $0.020 \pm 0.005$ & $-0.022 \pm 0.005$ & $0.043 \pm 0.011$ & $-0.043 \pm 0.011$ & $0.049 \pm 0.032$ & $-0.060 \pm 0.032$ \\
    \hline
    \end{tabular}
    }
    \caption{Summary of \textit{IXPE} observations of Cygnus X-1. For each bin we list the orbital phase $\phi$ (at bin midpoint computed using the \citet{Brocksopp1999} ephemeris), hardness ratio $\mathrm{HR}=(4\text{--}8\,\mathrm{keV})/(2\text{--}4\,\mathrm{keV})$, and normalized Stokes parameters $q$ and $u$ with $1\sigma$ uncertainties in three energy bands. ObsID 01002901 was processed with energy-scale corrected Level-2 event files (see Section~\ref{sec:2.Data}). All 26 bins from 12 observation IDs are included.}
    \label{tab:observations}
\end{table*}

\section{Spectral State Dependence} \label{sec:3.Dependence}
As shown in Figure~\ref{fig:1_PDPAvsMJD}, the PD is higher in the hard state than in the soft state. In the following, we first characterize the spectral state dependence of polarization before we search for geometric orbital signals.

\subsection{Correlation between Stokes Parameters and HR}
\begin{figure*}
    \hspace{-1cm}
    \centering
    \includegraphics[width=\textwidth]{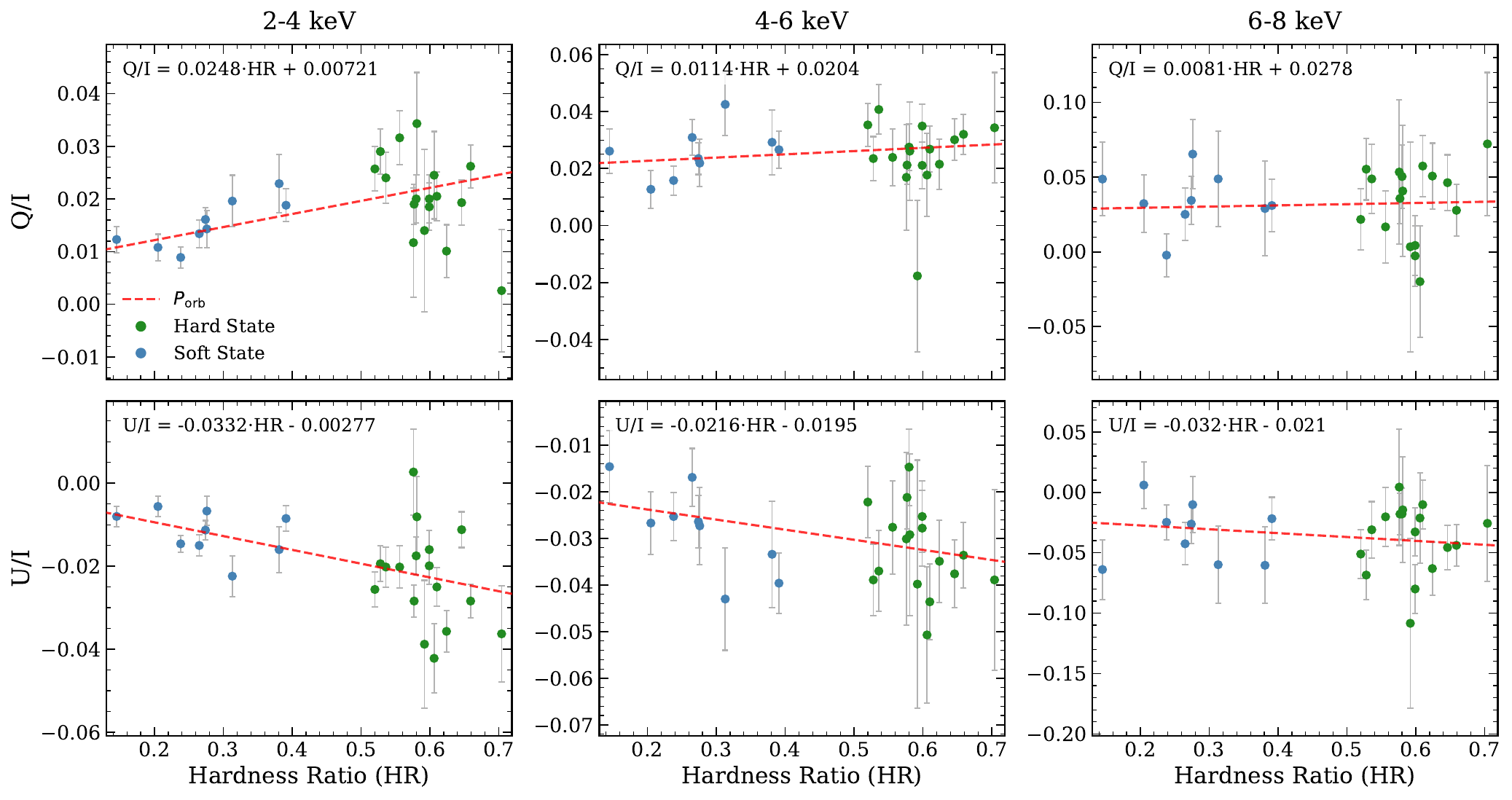}
    \caption{Normalized Stokes parameters $q$ (top row) and $u$ (bottom row) as a function of hardness ratio for the 2--4\,keV (left), 4--6\,keV (middle), and 6--8\,keV (right) energy bands.
    The dashed lines show the HR-dependent trends extracted from the first harmonic regression ($P_{\rm orb}$; Section~\ref{sec:4.HarmReg}).
    Full fit parameters for both models are provided in Table~\ref{tab:hr_fits}.}
    \label{fig:3_QUvsHR}
\end{figure*}

Figure~\ref{fig:3_QUvsHR} shows linear correlations between Stokes parameters ($q$, $u$) and the HR across all three energy bands.
As the source spectrum hardens, $q$ becomes more positive while $u$ becomes more negative.
This reflects the growing contribution of Comptonized coronal emission compared to the relatively unpolarized thermal disk emission \citep{Schnittman2009, Steiner2024}.
The dashed lines indicate the HR-dependent trends extracted from the simultaneous harmonic regression assuming the first harmonic orbital period ($P_{\rm orb}$; detailed in Section~\ref{sec:4.HarmReg}).
The slopes (e.g., $m_q \approx 0.025$ at 2--4\,keV; Table~\ref{tab:hr_fits}) imply that minor spectral fluctuations can shift polarization by amounts comparable to measurement precision, motivating a simultaneous spectral and orbital modeling approach.

\begin{table*}[ht]
    \centering
    \hspace{-1.5cm}
    \begin{tabular}{lc|cc|cc}
        \hline\hline
         & & \multicolumn{2}{c|}{$P_{\rm orb}$ Model ($n=1$)} & 
             \multicolumn{2}{c}{$P_{\rm orb}/2$ Model ($n=2$)} \\
        Band & Parameter & Slope ($m$) & Intercept ($b$) & 
                           Slope ($m$) & Intercept ($b$) \\
        \hline
        2--4\,keV & $q$ & $\phantom{-}0.0248 \pm 0.0044$ & $\phantom{-}0.0072 \pm 0.0018$ & 
                          $\phantom{-}0.0245 \pm 0.0045$ & $\phantom{-}0.0079 \pm 0.0019$ \\
                  & $u$ & $-0.0332 \pm 0.0044$           & $-0.0028 \pm 0.0018$           & 
                          $-0.0306 \pm 0.0045$           & $-0.0044 \pm 0.0019$ \\
        \noalign{\smallskip}
        4--6\,keV & $q$ & $\phantom{-}0.0115 \pm 0.0095$ & $\phantom{-}0.0204 \pm 0.0044$ & 
                          $\phantom{-}0.0134 \pm 0.0096$ & $\phantom{-}0.0207 \pm 0.0047$ \\
                  & $u$ & $-0.0217 \pm 0.0095$           & $-0.0195 \pm 0.0044$           & 
                          $-0.0231 \pm 0.0096$           & $-0.0171 \pm 0.0047$ \\
        \noalign{\smallskip}
        6--8\,keV & $q$ & $\phantom{-}0.0081 \pm 0.0259$ & $\phantom{-}0.0278 \pm 0.0125$ & 
                          $\phantom{-}0.0300 \pm 0.0262$ & $\phantom{-}0.0154 \pm 0.0132$ \\
                  & $u$ & $-0.0319 \pm 0.0259$           & $-0.0210 \pm 0.0125$           & 
                          $-0.0160 \pm 0.0262$           & $-0.0294 \pm 0.0132$ \\
        \hline
    \end{tabular}
    \caption{
    Best-fit $q(\textrm{HR})$ and $u(\textrm{HR})$ slopes and intercepts from the simultaneous harmonic regression (Equation~\ref{eq:simultaneous}).
    The table presents the HR-dependent parameters for both the first ($P_{\rm orb}$) and the second harmonic ($P_{\rm orb}/2$) models.
    Uncertainties are the nominal $1\sigma$ values from the weighted least-squares covariance matrix.
    }
    \label{tab:hr_fits}
\end{table*}

The HR is derived from the same photon counts used to calculate $q$ (or $u$), so a statistical covariance between them is expected in principle.
However, its magnitude is small.
Inverting the per-bin Stokes uncertainty $\sigma_q \simeq \sqrt{2}/(\mu\sqrt{N})$, with modulation factors $\mu \approx 0.2$ at 2--4\,keV and $\mu \approx 0.6$ at 6--8\,keV \citep{DiMarco2022}, implies $N \approx 2\times10^6$ counts per bin in the softest band and $N \approx 1\times10^4$ in the hardest.
We simulated Poisson fluctuations at these count levels for a source of constant polarization, so that any correlation between HR and $q$ arises purely from the shared counts.
The induced correlation is $|r| < 0.004$ in both bands, far below the level at which either the fitted HR slopes or the recovered orbital amplitudes would be affected.

\subsection{Energy and State Dependence of Polarization}
\begin{figure*}
    \centering
    \hspace{-1cm}
    \includegraphics[width=\textwidth]{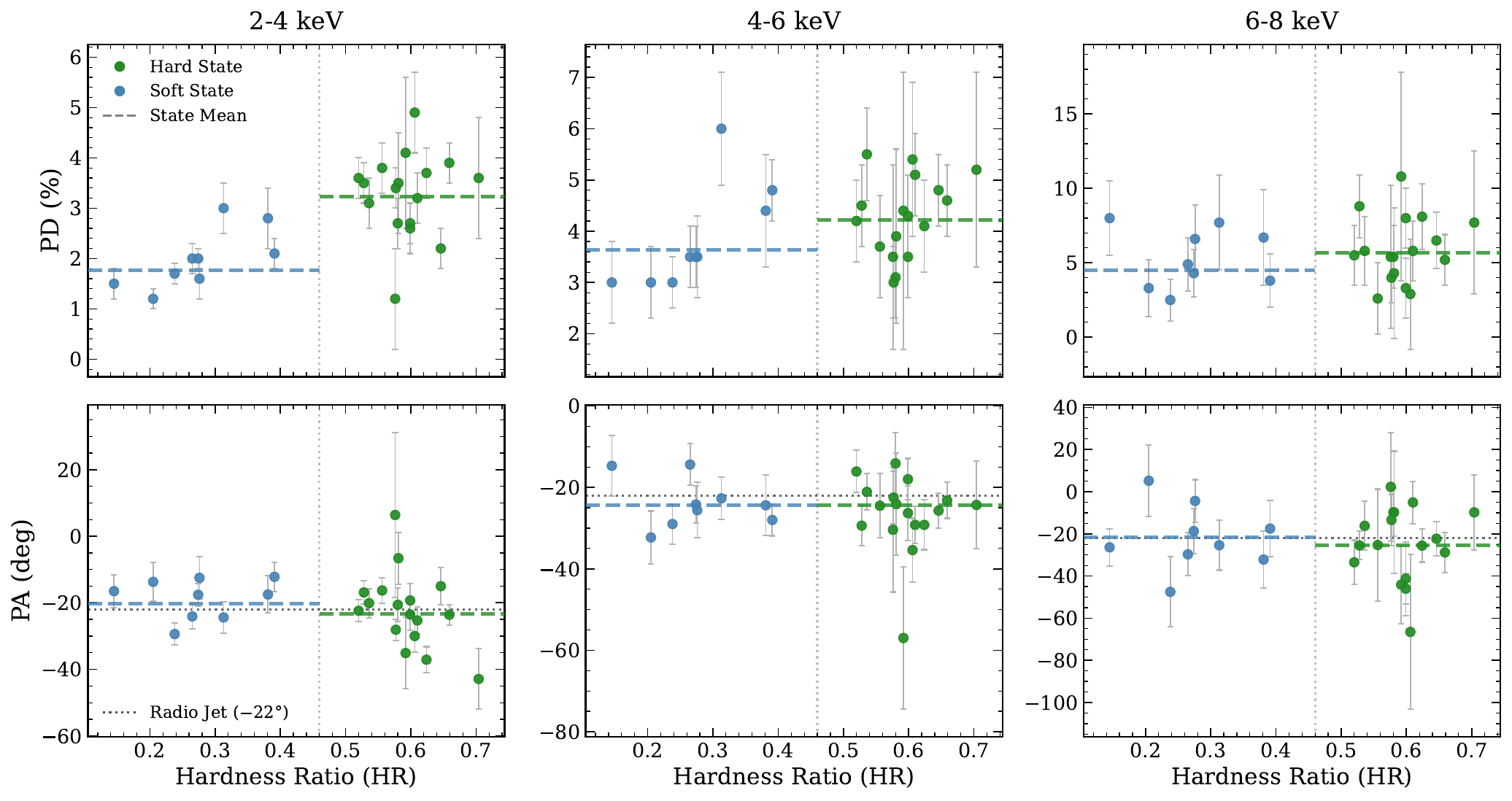}
    \caption{Polarization degree (PD, top row) and polarization angle (PA, bottom row) as a function of the hardness ratio (HR) in the 2--4, 4--6, and 6--8\,keV energy band columns (left to right). Green and blue points represent observations in the hard ($\mathrm{HR} \gtrsim 0.46$) and soft ($\mathrm{HR} \lesssim 0.46$) states, respectively. The vertical dotted line shows the approximate state transition boundary. The horizontal colored dashed lines indicate the mean polarization values for each state, and the black dotted line in the PA panels represents the position angle of the radio jet.}
    \label{fig:4_PDPAvsHR_3x2}
\end{figure*}

\begin{table}[ht]
    \centering
    \begin{tabular}{llccc}
    \hline\hline
    Band & State & $N$ & $\langle\mathrm{PD}\rangle$ & $\langle\mathrm{PA}\rangle$ \\
    \hline
    2--4\,keV   & Soft &  9 & 1.77\% $\pm$ 0.09\% & $-20.3^\circ \pm 1.5^\circ$ \\
                & Hard & 17 & 3.23\% $\pm$ 0.12\% & $-23.4^\circ \pm 1.1^\circ$ \\
    \noalign{\smallskip}
    4--6\,keV   & Soft &  9 & 3.64\% $\pm$ 0.23\% & $-24.4^\circ \pm 1.8^\circ$ \\
                & Hard & 17 & 4.22\% $\pm$ 0.22\% & $-24.4^\circ \pm 1.5^\circ$ \\
    \noalign{\smallskip}
    6--8\,keV   & Soft &  9 & 4.48\% $\pm$ 0.65\% & $-21.7^\circ \pm 3.9^\circ$ \\
                & Hard & 17 & 5.68\% $\pm$ 0.56\% & $-25.5^\circ \pm 2.7^\circ$ \\
    \hline
    \end{tabular}
    \caption{Weighted mean polarization degree ($\langle\mathrm{PD}\rangle$) and angle ($\langle\mathrm{PA}\rangle$) per energy band and spectral state, along with their corresponding $1\sigma$ errors of the weighted mean. Note that $\langle\mathrm{PD}\rangle$ values reported here are computed from the weighted means of PD directly and therefore retain its known positive bias \citep{Clarke1983}, which is most significant in the 6--8\,keV band where the per-bin signal-to-noise is lowest. All statistical tests in this paper are instead performed on the normalized Stokes parameters ($q,u$), which are unbiased.}
    \label{tab:pd_pa_mean}
\end{table}

Figure~\ref{fig:4_PDPAvsHR_3x2} illustrates the corresponding PD and PA measurements as a function of the HR for the three energy bands. To quantify these, we summarize the inverse-variance weighted mean PD and PA ($w_i=1/\sigma_i^2$) in each energy band, separated by spectral state, in Table~\ref{tab:pd_pa_mean}. We observe three main trends:
\begin{enumerate}
    \item PD increases with energy in all states, rising from $\sim$1.8--3.2\% at 2--4\,keV to $\sim$4.5--5.7\% at 6--8\,keV. This energy dependence can be explained in different ways for the soft state and hard state. In both states, reflected emission becomes more dominant at higher energies \citep{Steiner2024}. Furthermore, the harder emission is more strongly dominated by coronal emission, and the polarization of the coronal emission increases with energy as higher-energy photons scatter more often than low-energy photons.
    \item As shown by the mean dashed lines in the left column of Figure~\ref{fig:4_PDPAvsHR_3x2}, the hard-to-soft transition is most pronounced at 2--4\,keV, where unpolarized thermal disk emission dominates the flux in the soft state. At higher energies, the difference between the two states narrows, with the 4--6\,keV band showing the smallest absolute difference (3.64\% vs.\ 4.22\%).
    \item The PA remains stable across all bands and states. This angle is closely aligned with the radio jet axis at $\approx -22^\circ$ \citep{MillerJones2021}, with our measured weighted means constrained between $-20^\circ$ and $-26^\circ$ across all bands and states.
\end{enumerate}

\section{Model-Independent Search for Orbital Modulation} \label{sec:4.HarmReg}
Having characterized the dependence of polarization on the spectral state, we now search for a geometric signal modulated by the orbital phase. The fundamental challenge is that without simultaneously modeling both the $\mathrm{HR}$ and $\phi$, stochastic spectral variations could be misidentified as orbital modulation, or an orbital signal could be masked within the spectral variations.

\subsection{Rationale for Simultaneous Regression}
To decouple these effects, we must verify that the spectral state and orbital phase are independent. If the $\mathrm{HR}$ were systematically higher at certain orbital phases, a regression algorithm could not uniquely attribute polarization modulation to a specific predictor (a statistical issue known as multicollinearity) and would inflate parameter uncertainties.

Figure~\ref{fig:2_HRvsOP} demonstrates that both hard (HR $ \gtrsim 0.46$) and soft (HR $\lesssim 0.46$) states are evenly distributed across the full range of orbital phases. 
While Cygnus X-1 shows strong orbital modulation in X-ray absorption, the HR shows no significant correlation with orbital phase in our data set.
This is consistent with absorption dips being short-lived and concentrated near $\phi \approx 0$ \citep{Hanke2009}, such that $\sim$1-day binning smears them out, and with accretion-state-driven continuum changes dominating the HR dynamic range.
We quantitatively verified this independence by computing the Pearson correlation coefficient ($r$) between the HR and the orbital harmonics, defined as:
\begin{equation}
    r = \frac{\sum_{i} (X_i - \bar{X})(Y_i - \bar{Y})}{\sqrt{\sum_{i} (X_i - \bar{X})^2 \sum_{i} (Y_i - \bar{Y})^2}},
\end{equation}
where $X$ represents the HR and $Y$ represents the orbital harmonic terms.
Both quantities are evaluated as they are being regressed, using the window-averaged basis functions of Equation~\ref{eq:basis}.
For the first harmonic, we found no statistically significant linear correlation between the HR and the orbital phase ($r = 0.15$, $p = 0.47$ for $\mathcal{C}^{(1)}$; $r = -0.001$, $p = 0.99$ for $\mathcal{S}^{(1)}$).
Similarly, for the second harmonic, the correlation remains statistically insignificant ($r = 0.29$, $p = 0.15$ for $\mathcal{C}^{(2)}$; $r = 0.26$, $p = 0.21$ for $\mathcal{S}^{(2)}$).

Furthermore, we checked for multicollinearity among the regression covariates using the variance inflation factor (VIF), defined for a given covariate $j$ as:
\begin{equation}
    \mathrm{VIF}_j = \frac{1}{1 - R_j^2},
\end{equation}
where $R_j^2$ is the coefficient of determination obtained by regressing covariate $j$ against all other covariates.
A VIF near 1 indicates no collinearity, whereas values exceeding 5--10 indicate that uncertainties are highly inflated \citep{Obrien2007}.
In our 1st harmonic model, the VIFs are $1.02$ for HR, $1.11$ for $\mathcal{C}^{(1)}$, and $1.08$ for $\mathcal{S}^{(1)}$.
For the 2nd harmonic model, the VIFs are $1.16$ for HR, $1.09$ for $\mathcal{C}^{(2)}$, and $1.07$ for $\mathcal{S}^{(2)}$.
Across both models, all covariates yield $\mathrm{VIF} < 1.16$, confirming that the spectral and orbital terms behave as independent predictors and supporting the use of a simultaneous regression approach.

\subsection{Simultaneous Harmonic Regression Model}
Periodicity searches in astrophysics often include epoch folding \citep{Leahy1987} or Lomb-Scargle periodograms \citep{Lomb1976, Scargle1982}, which are designed for univariate time series and do not accommodate a continuous covariate such as the HR. Instead of employing sequential subtraction, which can overcorrect the pure orbital signatures, we perform simultaneous regression. We probe different possible modulation timescales within the binary system by fitting for both the orbital period ($P_{\rm orb}$, $n=1$) and the half-orbital period ($P_{\rm orb}/2$, $n=2$) components. For each harmonic order $n$, we parameterize the orbital dependence with Fourier components, $A\cos(2n\pi\phi) + B\sin(2n\pi\phi)$, which can represent a sinusoidal oscillation of arbitrary amplitude and phase offset with two linear coefficients.

This stacked analysis implicitly assumes that any orbital modulation is stationary in amplitude and phase across the 2022--2024 campaign and across spectral states.
A transient or epoch-dependent signal would be diluted by averaging over the full dataset (see Section~\ref{sec:6.Discussion} for further discussion of this assumption).

We fit $q$ and $u$ simultaneously, modeling each for a given harmonic $n$ as:
\begin{equation} \label{eq:simultaneous}
    S_i = \beta_0 + \beta_{\rm HR} \cdot \mathrm{HR}_i +
    A \cos(2n\pi \phi_i) + B \sin(2n\pi \phi_i) + \epsilon_i,
\end{equation}
where $S \in \{q, u\}$, the coefficients $A$ and $B$ parameterize the orbital modulation at the specified frequency, and $\epsilon_i$ is the measurement noise.
Rather than evaluating the harmonic basis at the bin midpoint, we average it over the phase window spanned by each bin,
\begin{equation}\label{eq:basis}
    \begin{split}
    \mathcal{C}^{(n)}_i
        &\equiv \frac{1}{\Delta\phi_i} \int_{\phi_{{\rm min},i}}^{\phi_{{\rm max},i}} \!\! \cos(2\pi n \phi)\, d\phi \\
        &= \frac{\sin(2\pi n \phi_{{\rm max},i}) - \sin(2\pi n \phi_{{\rm min},i})} {2\pi n \, \Delta\phi_i},
    \end{split}
\end{equation}
with $\Delta\phi_i = \phi_{{\rm max},i} - \phi_{{\rm min},i}$ and $\mathcal{S}^{(n)}_i$ defined analogously for $\sin(2\pi n \phi)$.
The best-fit parameters are found by minimizing the $\chi^2$ statistic:
\begin{equation}\label{eq:chi2}
    \chi^2 = \sum_i \left[
    \frac{(q_i - \hat{q}_i)^2}{\sigma_{q,i}^2} +
    \frac{(u_i - \hat{u}_i)^2}{\sigma_{u,i}^2}
    \right],
\end{equation}
where $\hat{q}_i$ and $\hat{u}_i$ are the model predictions from Equation~\ref{eq:simultaneous}, and $\sigma_{q,i}$ and $\sigma_{u,i}$ are the $1\sigma$ measurement uncertainties. This yields eight free parameters in total per harmonic configuration (intercept $\beta_0$, slope $\beta_{\rm HR}$, and harmonic coefficients $A, B$ for each Stokes parameter).

Because Equation~\ref{eq:basis} integrates over each bin's actual phase window, the fitted coefficients $A$ and $B$ are intrinsic amplitudes by construction, and no post correction for finite-bin smearing is required.
The attenuation absorbed into the basis is quantified by $|W^{(n)}_i| = [(\mathcal{C}^{(n)}_i)^2 + (\mathcal{S}^{(n)}_i)^2]^{1/2}$, which ranges from 0.948 to 1.000 for $n=1$ and from 0.803 to 0.999 for $n=2$ across our 26 bins.

Equation~\ref{eq:basis} assumes that the exposure is continuous across each bin, whereas in practice it is interrupted by Earth occultation.
We therefore recomputed the basis two further ways: averaging only over the good time intervals recorded in the event files and over the arrival times of the source events themselves, which weights each moment by the count rate in the same way $q$ and $u$ are.
Relative to the uniform-window basis, the individual $|W^{(2)}_i|$ shift by at most 0.16, extending the range to 0.757--0.999, but the fitted amplitudes and permutation $p$-values are unchanged to within 0.02 percentage points and 0.005 respectively in every band and harmonic.
We therefore retain the closed-form window average of Equation~\ref{eq:basis}, which is analytic and requires no event-level information.

We then define the total intrinsic orbital modulation amplitude for the $n$-th harmonic as:
\begin{equation} \label{eq:Rorb}
    R = \sqrt{(A_q^2 + B_q^2 + A_u^2 + B_u^2)/2}.
\end{equation}
Since linear polarization is characterized by the normalized Stokes parameters, orbital modulation can be inferred from changes in $q$ or $u$.
Each pair $(A_q, B_q)$ and $(A_u, B_u)$ describes a sinusoidal oscillation whose peak amplitude is $\sqrt{A^2+B^2}$, for $q$ and $u$ respectively.
$R$ is defined as the root mean square (RMS) of these two peak amplitudes (hence the division by two inside the square root of Equation~\ref{eq:Rorb}), characterizing the overall amplitude of the combined ($q, u$) modulation.
This modulation reflects the geometric scattering contribution originating from the companion's stellar wind or intrabinary structures.
We refer to this quantity computed from the observed data as $R_{\rm obs}$, from phase-shuffled permutation trials as $R_{\rm null}$, and from injection-recovery trials as $R_{\rm inj}$ (injected) and $R_{\rm rec}$ (recovered).
We use three amplitude conventions. \citet{VanderMeulen2026} quote \emph{minimum-to-maximum} PD amplitudes of 0.25, 0.81, and 1.24 percentage points, the \emph{peak} amplitude of a single Stokes parameter is $\sqrt{A^2+B^2}$, and the \emph{RMS} amplitude $R$ of Equation~\ref{eq:Rorb} is the quantity we constrain. Because the \texttt{SKIRT} templates are not single sinusoids, we compute the predicted RMS amplitudes of $0.10\%$, $0.33\%$, and $0.49\%$ directly from the mean-subtracted templates rather than converting from the minimum-to-maximum figures. The second harmonic gives $0.06\%$, $0.27\%$, and $0.41\%$.

\subsection{Model Validation via Injection-Recovery}
\label{sec:4.3}
Before interpreting $R_{\rm obs}$, we must validate the accuracy and potential biases of our simultaneous regression model.
Because $R_{\rm obs}$ is a positive-definite quantity, even in the absence of a true physical orbital signal, random noise may cause the model to fit small, non-zero values for $A$ and $B$, resulting in a positive $R_{\rm obs}$.
This statistical ``noise floor'' defines the minimum detectable amplitude for our dataset.

To quantify this noise floor, we ran injection-recovery simulations independently for both the $n=1$ and $n=2$ harmonic models.
Injection-recovery is a standard tool for characterizing the response and bias of an amplitude estimator in the low signal-to-noise regime.
For each trial, a fake orbital signal with known amplitude $R_{\rm inj}$ was injected into  simulated pure noise based on the observational uncertainties by adding independent harmonic components:
\begin{equation} \label{eq:injection}
    S_i^{\rm inj} = \mathcal{N}(0, \sigma_{S,i}^2) + A_S \mathcal{C}^{(n)}_i + B_S \mathcal{S}^{(n)}_i,
\end{equation}
where $S \in \{q, u\}$ denotes the normalized Stokes parameters, and $\mathcal{N}(0, \sigma_{S,i}^2)$ represents random Gaussian noise drawn from their respective measurement errors.
The four coefficients ($A_q, B_q, A_u, B_u$) are drawn independently from a standard normal distribution and then normalized such that $R_{\rm inj} = \sqrt{(A_q^2 + B_q^2 + A_u^2 + B_u^2)/2}$.
This procedure was repeated for $10^4$ trials at each of 100 injected amplitudes spanning $0.0\%$--$3.5\%$.
The full distribution of $R_{\rm rec}$ is retained at each grid point, since its percentiles are used to derive upper limits in Section~\ref{sec:4.6.UL}.

\begin{figure*}[ht]
    \centering
    \includegraphics[width=\linewidth]{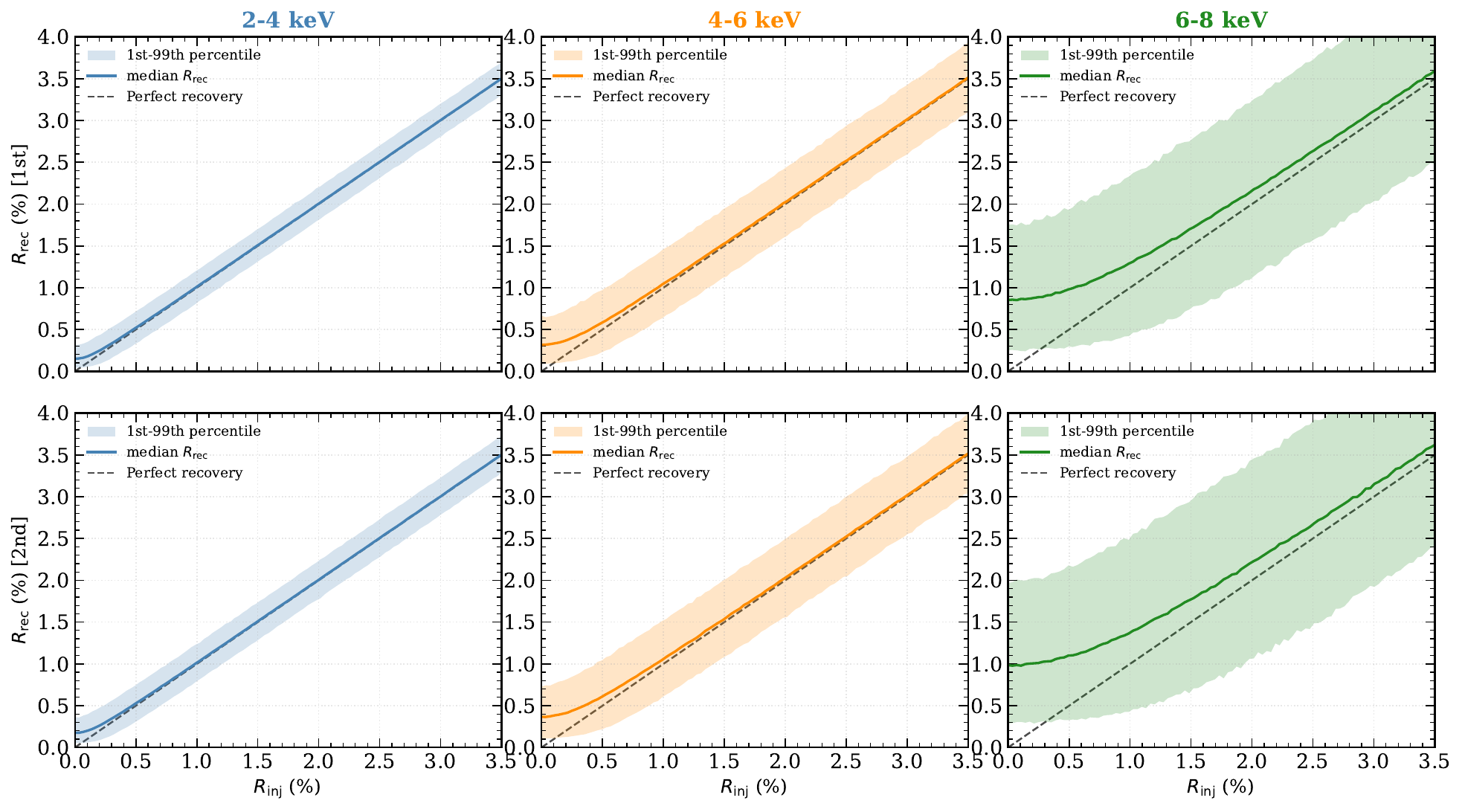} 
    \caption{Injection-recovery test for all three energy bands (columns) under the 1st (top row) and 2nd (bottom row) harmonic models.
    For each injected amplitude $R_{\rm inj}$ (x-axis), $10^4$ trials were run with fake signals injected into simulated noise drawn from the observational uncertainties (Equation~\ref{eq:injection}).
    Solid curves show the median recovered amplitude $R_{\rm rec}$ and shaded bands the 1st--99th percentile range.
    The black dashed line indicates perfect recovery ($R_{\rm rec} = R_{\rm inj}$).
    At low $R_{\rm inj}$, $R_{\rm rec}$ plateaus at a non-zero noise floor (Section~\ref{sec:4.HarmReg}), indicating that a true signal with an amplitude below this floor cannot be reliably recovered.
    Above this floor, recovery is approximately linear.
    The lower edge of each band is the 1st percentile curve used to derive the upper limits of Section~\ref{sec:4.6.UL}.}
    \label{fig:5_injection}
\end{figure*}

Figure~\ref{fig:5_injection} shows the results of these simulations.
At $R_{\rm inj} = 0.0\%$, the recovered amplitude converges to a noise-floor of $\sim$0.16\%, $\sim$0.33\%, and $\sim$0.88\% in the 2--4, 4--6, and 6--8\,keV bands, respectively, for the first harmonic, and to $\sim$0.18\%, $\sim$0.38\%, and $\sim$1.01\% for the second harmonic.
The second-harmonic floors are systematically higher because the finite bin width attenuates the $n=2$ signal more strongly (0.803--0.999 vs. 0.948--1.000 for $n=1$), so a given intrinsic modulation appears weaker in the data.
At the maximum injected amplitude of $R_{\rm inj} = 3.50\%$, the mean recovered amplitudes are $3.50\%$, $3.51\%$, and $3.60\%$ for the first harmonic, and $3.50\%$, $3.52\%$, and $3.62\%$ for the second harmonic.
Well above their respective noise floors, $R_{\rm rec}$ increases monotonically and approximately linearly with $R_{\rm inj}$, following the injected amplitude with only a small offset that is largest in the noise-dominated 6--8\,keV band.
This confirms that the simultaneous regression successfully recovers an injected signal at both frequencies without significant absorption into the HR trend.
This holds because HR and the orbital harmonics are uncorrelated in the present dataset (VIF $< 1.16$).
In general, an orbital signal correlated with HR variations could be partially absorbed by the spectral term.
All simulations use explicitly seeded random number generators, so every quoted permutation and injection-recovery value is exactly reproducible.

\subsection{Permutation Test for Significance} \label{sec:4.4}
To evaluate whether the observed amplitudes ($R_{\rm obs}$) represent a statistically significant signal, we compare them against a null distribution representing the expected amplitudes in the absence of any true orbital modulation.
Standard parametric tests, such as the F-test for nested models, assume that residuals are normally distributed.
However, given our limited degrees of freedom (52 $(q,u)$ measurements across 26 phase bins and 8 free parameters), which overstates the effective number of independent measurements in the 2--4\,keV band where residuals are correlated within observations (Appendix~\ref{sec:appendixB}), and the positive-definite, non-Gaussian nature of our test statistic ($R_{\rm obs}$), parametric F-tests can be unreliable.
We therefore adopt a non-parametric permutation test \citep{Good2005}, which makes no underlying assumptions about the distribution of the test statistics.
The permutation test is performed separately for the first and second harmonics as follows:
\begin{enumerate}
    \item For each of $10^4$ permutation trials, the orbital phase assignments $\{\phi_i\}$ are randomly shuffled among the 26 bins, while the $(q_i, u_i, \mathrm{HR}_i, \sigma_{q, i}, \sigma_{u, i})$ tuples are held fixed.
    \item The simultaneous regression (Equation~\ref{eq:simultaneous}) is re-fit to each shuffled dataset, and the resulting amplitude $R_{\rm null}$ is recorded.
    \item The $p$-value is computed as the fraction of permutations for which $R_{\rm null} \geq R_{\rm obs}$.
\end{enumerate}
This test destroys any orbital periodicity while preserving the spectral correlations.
Note that the phase assignments are permuted as $(\phi_{\rm min}, \phi_{\rm max})$ pairs, so that each shuffled dataset retains the true distribution of bin durations.
The 99th percentile of the $R_{\rm null}$ distribution is the amplitude that would be exceeded in 1\% of trials under the null hypothesis.
We refer to it as the 99\% detection threshold, which is distinct from an upper limit on the true modulation amplitude, a distinction we return to in Section~\ref{sec:4.6.UL}.

We note that instrumental effects, such as residual spurious modulation from imperfect in-flight calibration, could in principle mimic a low-level orbital signal.
However, because such instrumental signals are stationary in detector coordinates and not phase-locked to the 5.6-day binary period, they are absorbed into the fitted intercept $\beta_0$, and our permutation test treats any such offset as part of the null.

Our criterion of $p < 0.01$ for a secure detection is set by the number of tests.
With three energy bands and two harmonics, a Bonferroni correction of a nominal 5\% level gives $0.05/6 = 0.0083$, which we round to $0.01$.
We also computed the family-wise $p$-value by shuffling the phases $2\times10^4$ times and recomputing all six statistics from each shuffle.
Our smallest marginal value, in the 6--8\,keV second harmonic, is exceeded in 23.4\% of permutations, so $p_{\rm FW} = 0.23$.
Thus, none of the six results is significant once the number of tests is taken into account.

\subsection{Simultaneous Harmonic Regression Results} \label{sec:4.5.Results}
\begin{figure*}
    \centering
    \includegraphics[width=\linewidth]{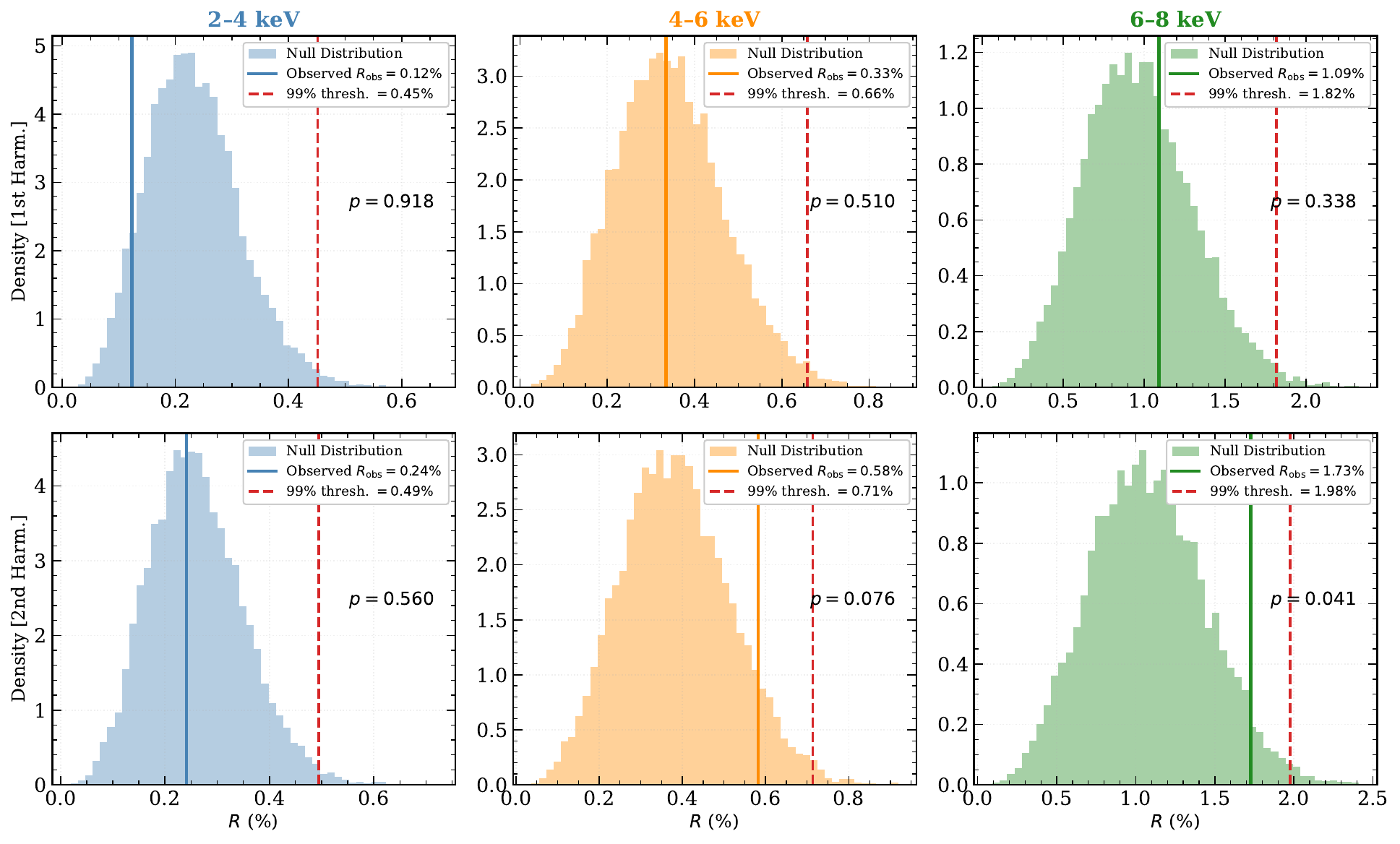}
    \caption{Null amplitude distributions from $10^4$ phase-shuffled datasets for the 2--4\,keV (blue), 4--6\,keV (orange), and 6--8\,keV (green) bands for the first (upper) and second (lower) harmonics. The vertical solid line marks the observed amplitude $R_{\rm obs}$, the quantity of interest, and the dashed line marks the 99th percentile detection threshold. A signal would appear as $R_{\rm obs}$ lying in the upper tail of the null distribution.}
    \label{fig:6_PermutationNull}
\end{figure*}

Applying the simultaneous harmonic regression across all three bands, we find that the recovered amplitudes ($R_{\rm obs}$) are consistent with the permutation null distributions, indicating no statistically significant orbital modulation.
The resulting $p$-values and detection thresholds for both first and second harmonic periods are summarized in Table~\ref{tab:upper_limits} and Figure~\ref{fig:6_PermutationNull}.

\begin{table*}[ht]
    \centering
    \hspace{-1.5cm}
    \begin{tabular}{lccccc}
        \hline
        Band & Harmonic & $R_{\rm obs}$ (\%) & $p$-value & 99\% thresh. (\%) & 99\% UL (\%) \\
        \hline
        \multirow{2}{*}{2--4\,keV} 
        & 1st ($P_{\rm orb}$)   & 0.12 & 0.918 & 0.45 & 0.25 \\
        & 2nd ($P_{\rm orb}/2$) & 0.24 & 0.560 & 0.49 & 0.42 \\
        \hline
        \multirow{2}{*}{4--6\,keV} 
        & 1st ($P_{\rm orb}$)   & 0.33 & 0.510 & 0.66 & 0.63 \\
        & 2nd ($P_{\rm orb}/2$) & 0.58 & 0.076 & 0.71 & 0.97 \\
        \hline
        \multirow{2}{*}{6--8\,keV} 
        & 1st ($P_{\rm orb}$)   & 1.09 & 0.338 & 1.82 & 1.96 \\
        & 2nd ($P_{\rm orb}/2$) & 1.73 & 0.041 & 1.98 & 2.78 \\
        \hline
    \end{tabular}
    \caption{Results of the permutation test.
    The harmonic basis is averaged over each bin's actual phase window (Equation~\ref{eq:basis}).
    The 99\% threshold column is the 99th percentile of the phase-shuffled null distribution, i.e.\ the amplitude required to reject the null hypothesis at 99\% confidence.
    The final column is the 99\% confidence upper limit on the true modulation amplitude, derived from the injection-recovery distributions (Section~\ref{sec:4.6.UL}).
    Note that these two are distinct quantities.}
    \label{tab:upper_limits}
\end{table*}

For the first harmonic ($P_{\rm orb}$), the $p$-values range from 0.338 to 0.918, meaning the observed amplitudes are statistically indistinguishable from the permutation null distribution.
The corresponding 99\% detection thresholds are 0.45\%, 0.66\%, and 1.82\% for the 2--4, 4--6, and 6--8\,keV bands, respectively.
Similarly, for the second harmonic ($P_{\rm orb}/2$), the test shows $p$-values of 0.560, 0.076, and 0.041, with thresholds of 0.49\%, 0.71\%, and 1.98\%.
None of these values reaches the threshold for a statistically secure detection ($p < 0.01$).
Upper limits on the true amplitude are derived separately in Section~\ref{sec:4.6.UL}.
The tightest constraint is obtained in the 2--4\,keV band for both harmonics, which is also where the stellar companion and wind scattering model predicts the smallest modulation amplitude \citep{VanderMeulen2026}.
The 6--8\,keV second harmonic yields the smallest $p$-value in the dataset ($p = 0.041$, $R_{\rm obs} = 1.73\%$ against a threshold of $1.98\%$).
Accounting for the six tests performed, the family-wise $p$-value is 0.23 (Section~\ref{sec:4.4}): a smallest value this low arises by chance in 23\% of phase-shuffled trials. Together with the noise floor $\sim$1.01\% in this band, we regard it as consistent with a noise-dominated non-detection.

\subsection{Upper Limits on the Modulation Amplitude} \label{sec:4.6.UL}
The 99th percentile of the permutation null distribution is the amplitude that noise alone would produce in 1\% of phase-shuffled trials.
This sets the bar for claiming a detection, but it does not tell us how large the true modulation could have been.
Because $R$ is a positive-definite quantity, noise alone still yields a positive value as shown by the recovered amplitude plateauing at the noise floors of Section~\ref{sec:4.3} (e.g., 0.18\% at 2--4\,keV for $n=2$) rather than at zero.
The measured $R_{\rm obs}$ therefore reads high when the signal is weak and cannot be converted into a limit.
We instead derive upper limits from the injection-recovery distributions.

For each injected amplitude $R_{\rm inj}$ we record the full distribution of recovered amplitudes over $10^4$ trials, on a grid of 100 values spanning $0$--$3.5\%$.
A given $R_{\rm inj}$ is excluded at 99\% confidence if fewer than 1\% of its trials yield $R_{\rm rec} \leq R_{\rm obs}$.
The 99\% upper limit is the smallest such $R_{\rm inj}$, or equivalently, it is where the 1st percentile curve of Figure~\ref{fig:5_injection} (the lower edge of the shaded band) crosses $R_{\rm obs}$.
The resulting limits are listed in the final column of Table~\ref{tab:upper_limits}: 0.25\%, 0.63\%, and 1.96\% at $P_{\rm orb}$, and 0.42\%, 0.97\%, and 2.78\% at $P_{\rm orb}/2$, for the 2--4, 4--6, and 6--8\,keV bands respectively.

\section{Model-Dependent Constraints: Comparison with 3D Radiative Transfer Simulations} \label{sec:5.BertModel}
To complement our model-independent harmonic search, we compare the observational data directly against 3D Monte Carlo radiative transfer predictions for polarization variations induced by stellar companion and wind scattering \citep{VanderMeulen2026}.
This model, computed using the \texttt{SKIRT} code \citep{VanderMeulen2023}, predicts minimum-to-maximum orbital modulation amplitudes of 0.25, 0.81, and 1.24 percentage points in the 2--4, 4--6, and 6--8\,keV bands, respectively.
The \texttt{SKIRT} templates are tabulated on a uniform grid in orbital phase and interpolated to the observed phases using periodic interpolation.
Consistent with the treatment of the harmonic basis (Equation~\ref{eq:basis}), the template is averaged over each bin's phase window rather than evaluated at the bin midpoint.

\begin{figure*}[t]
    \centering
    \includegraphics[width=\linewidth]{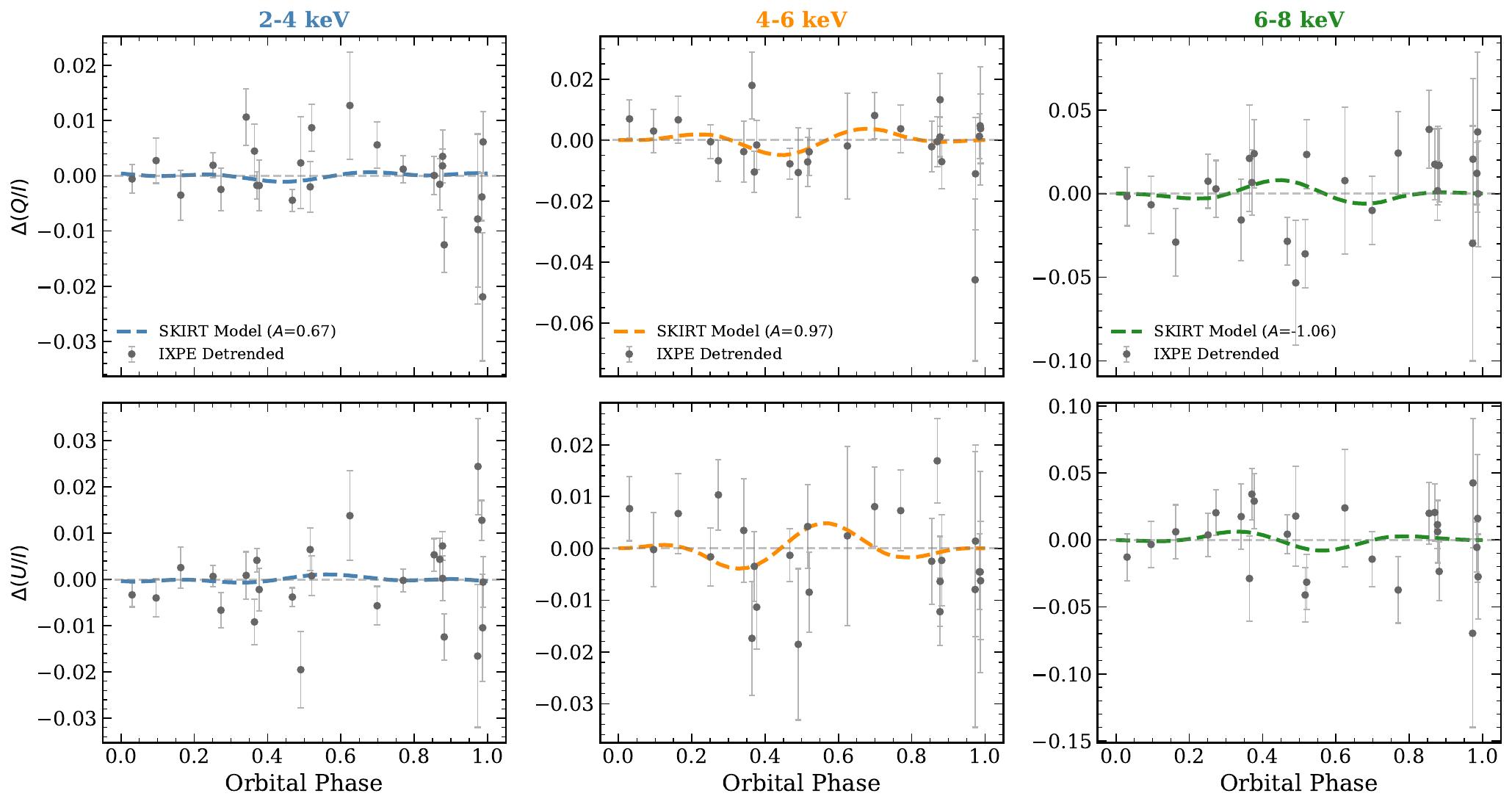}
    \caption{\textit{IXPE} Stokes residual curves ($q$ and $u$) phase-folded over the orbital period for the 2--4\,keV (left), 4--6\,keV (middle), and 6--8\,keV (right) energy bands. Black circles with error bars represent the observed data points after removing the HR trend. The colored curves show the best-fit 3D Monte Carlo radiative transfer templates predicted by the \texttt{SKIRT} simulation \citep{VanderMeulen2026}, scaled by the best-fit amplitude $A$ from Table~\ref{tab:bert_comparison}.}
    \label{fig:7_phase_folded}
\end{figure*}

We evaluate the agreement of the model's phase-resolved Stokes $Q/I$ and $U/I$ templates with our \textit{IXPE} observations by running a simultaneous $\chi^2$ fit.
The observed normalized Stokes parameters $S(\phi)$ as a function of the orbital phase $\phi$ are modeled as:
\begin{equation}
    S_{\rm model}(\phi)=b_{0,S}+b_{\rm HR,S}\cdot{\rm HR}+A\cdot SKIRT_S(\phi)
\end{equation}
where $S \in \{q, u\}$.
Here, ${\rm HR}$ is the observed hardness ratio from the \textit{IXPE}, and $SKIRT_S(\phi)$ is the mean-centered stellar companion and wind-scattering template predicted by the \texttt{SKIRT} simulation and interpolated at the corresponding orbital phase.
We use the \texttt{SKIRT} templates in the \textit{IXPE} reference frame (celestial north up), which adopt the observed clockwise sky-projected orbital motion of Cygnus X-1 and the same $\phi=0$ convention as the \citet{Brocksopp1999} ephemeris, so they compare directly to the \textit{IXPE} $(q,u)$ without further rotation or phase shift.
The \citet{Brocksopp1999} zero point, superior conjunction of the black hole, is the same orbital configuration as the inferior conjunction of the companion adopted by \citet{VanderMeulen2026}.
The free parameters determined by the fit include the baseline offsets ($b_{0,S}$), the linear HR dependency coefficients ($b_{\rm HR,S}$), and a single amplitude scaling factor $A$ applied simultaneously to both Stokes parameters.
In this model, $A=1$ indicates perfect agreement with the nominal \texttt{SKIRT} predictions, while $A=0$ indicates a complete absence of the predicted orbital modulation.
The results of these template fits are summarized in Table~\ref{tab:bert_comparison} and illustrated in Figure~\ref{fig:7_phase_folded}.

\begin{table*}[ht]
    \centering
    \hspace{-1.5cm}
    \begin{tabular}{lccc}
        \hline
        Band & Best-fit Amplitude ($A$) & $\Delta\chi^2$ for 1 d.o.f. & Significance \\
        \hline
        2--4\,keV & $\phantom{-}0.67 \pm 0.85$ & 0.62 & $0.8\sigma$ \\
        4--6\,keV & $\phantom{-}0.97 \pm 0.62$ & 2.46 & $1.6\sigma$ \\
        6--8\,keV & $-1.06           \pm 1.11$ & 0.90 & $0.9\sigma$ \\
        \hline
    \end{tabular}
    \caption{Direct comparison between the \textit{IXPE} data and the 3D stellar companion and wind-scattering model templates.
    The amplitude $A$ represents the scaling factor applied to the model, where $A=1$ represents the nominal prediction.
    The significance is derived from the $\Delta\chi^2$ improvement over a null model ($A=0$).
    The reduced $\chi^2$ of the fit is 1.98, 0.88, and 0.97 for the three bands.
    The excess in 2--4\,keV reflects the within-observation correlated scatter identified in Appendix~\ref{sec:appendixB}.
    Note that inflating $\sigma_A$ by $\sqrt{\chi^2_\nu} = 1.41$ in that band gives $A = 0.67 \pm 1.20$, so the conclusion is unchanged even with conservatively rescaled uncertainties.}
    \label{tab:bert_comparison}
\end{table*}

At all three energy bands, the data are consistent with no statistically significant orbital modulation above the measurement uncertainties, reflecting the limited sensitivity of the current dataset to the nominal model (Section~\ref{sec:6.Discussion}).
The best-fit amplitude parameters are $A = 0.67 \pm 0.85$, $A = 0.97 \pm 0.62$, and $A = -1.06 \pm 1.11$ at 2--4, 4--6, and 6--8\,keV, respectively.
Including the \texttt{SKIRT} template yields a maximum significance of $\sim1.6\sigma$ (in the 4--6\,keV band, where $\Delta\chi^2 = 2.46$).

While the stellar companion and wind-scattering model predicts an increasing modulation amplitude with energies, no statistically significant energy-dependent trend is detected 
in the current dataset.
The negative amplitude in the 6--8\,keV band is statistically consistent with zero ($0.9\sigma$) rather than suggesting anti-correlation with the model template.

\section{Discussion} \label{sec:6.Discussion}
\subsection{Comparison with Theoretical Predictions and Noise Limitations}
Our harmonic regression and \texttt{SKIRT} template fitting results do not reveal statistically significant $P_{\rm orb}$ or $P_{\rm orb}/2$ modulation in any of the three energy bands.
For the model-independent harmonic regression, the 99\% confidence upper limits on the $P_{\rm orb}$ modulation amplitude are 0.25\%, 0.63\%, and 1.96\% for the respective bands (0.42\%, 0.97\%, and 2.78\% at $P_{\rm orb}/2$).
For the \texttt{SKIRT} template fitting, the best-fit amplitude scaling factors are $A=0.67\pm0.85$, $A=0.97\pm0.62$, and $A=-1.06\pm1.11$, with the scaling factors in the soft and intermediate bands consistent with the nominal model ($A=1$) within 1$\sigma$ uncertainties.

These constraints do neither confirm nor refute the model because the predicted signal lies below our sensitivity.
As demonstrated by our injection-recovery simulations (Section~\ref{sec:4.HarmReg}), the positive-definite statistical noise floor of the current \textit{IXPE} dataset sits at $\sim0.18\%$, $\sim0.38\%$, and $\sim1.01\%$ for the three bands at $P_{\rm orb}/2$, the period at which the reflection signature is predicted, whereas the predicted stellar companion and wind-scattering signals correspond to RMS amplitudes of $\approx 0.10\%$, $\approx0.33\%$, and $\approx0.49\%$, computed directly from the \texttt{SKIRT} templates.
The predicted signal therefore lies below the noise floor in all three energy bands.
We note that $R_{\rm obs}$ from the $n=2$ regression measures, by construction, only the second-harmonic component of any modulation, whereas the RMS amplitudes quoted above are computed from the full templates.
Projecting the templates onto $n=2$ gives $0.06\%$, $0.27\%$, and $0.41\%$, so when both are restricted to the second harmonic, the predicted signal lies below the corresponding noise floors by factors of $\sim$3.2, $\sim$1.4, and $\sim$2.5.
Therefore, the lack of significant $P_{\rm orb}$ or $P_{\rm orb}/2$ signal is consistent with the stellar companion and  wind-scattering signatures remaining undetected rather than ruled out. 

\subsection{Sensitivity Limitations in the 6--8 keV Band}
The energy dependence of the predicted stellar companion and wind-scattering template is governed by the competing effects of photoelectric absorption and electron scattering \citep{Ahlberg2024, VanderMeulen2026} which are modulated by the binary geometry \citep[e.g.,][]{Brown1978, Kravtsov2020}.
At lower energies (2--4\,keV), soft X-rays are mostly absorbed in the stellar atmosphere of the companion star, suppressing the reflected signal \citep{Ahlberg2024, VanderMeulen2026}.
At higher energies, reduced photoelectric absorption allows X-rays to penetrate deeper into the focused stellar wind and reflect off the companion star surface, producing a larger net polarization modulation \citep{VanderMeulen2026}.

However, in our hardest energy band (6--8\,keV), the fit yields a negative scaling factor ($A = -1.06 \pm 1.11$).
While a negative amplitude would imply a modulation pattern out of phase with the model, its low statistical significance (0.9$\sigma$) is consistent with a null result ($A=0$).
Given the 0.9$\sigma$ significance, the negative sign carries no meaningful physical interpretation and reflects the dominant noise in this band.
The per-bin Stokes uncertainties in the 6--8\,keV band are roughly 3--4 times larger than in the 2--4\,keV band (Table~\ref{tab:observations}), a consequence of declining effective area of \textit{IXPE} at higher energies \citep{Weisskopf2022}.
The 6--8\,keV band is thus limited by photon statistics in the current dataset.

\subsection{Future Observations and Prospects}
Our observational constraints are noise-limited, and further observations are required to (i) achieve a secure detection of the nominal \texttt{SKIRT} model and (ii) constrain the underlying wind physics, using the template-fit formalism of Section~\ref{sec:5.BertModel}, which is linear and well-suited to this projection.

\textit{Exposure for a $5\sigma$ detection.} The current per-band uncertainties on the amplitude scaling factor are $\sigma_A = 0.85$, $0.62$, and $1.11$ for the 2--4, 4--6, and 6--8\,keV bands. If $A=1$ is the true value, the expected significances with the current dataset are $1.2\sigma$, $1.6\sigma$, and $0.9\sigma$, respectively, consistent with our measurements (Table~\ref{tab:bert_comparison}). Since $\sigma_A \propto 1/\sqrt{N}$ for $N$ comparable bins, a $5\sigma$ detection requires $\sigma_A \approx 0.2$. For the 4--6\,keV band alone, this corresponds to $(0.62/0.2)^2 \approx 10\times$ the current dataset ($\sim$260 bins). Combining all three bands in inverse-variance weighting gives $\sigma_{A,{\rm comb}} \approx 0.46$ at present (an expected $\sim 2.2\sigma$ for $A=1$). Reaching $5\sigma$ combined requires $(0.46/0.2)^2 \approx 5.3\times$ the current exposure ($\sim$140 bins). 
For reference, doubling the current exposure (as might be achieved with a single additional \textit{IXPE} observing campaign) would yield $\sigma_{A,{\rm comb}} \approx 0.33$, i.e., an expected $\sim 3\sigma$ combined, which is suggestive evidence, but not a secure detection.

\textit{Exposure to constrain the wind physics.} A $5\sigma$ detection is equivalent to a $\sim$20\% measurement of $A$, which in the single-scattering, optically thin limit translates approximately to a $\sim$20\% constraint on the wind density normalization \citep{Brown1978}, and would distinguish the nominal smooth-wind model ($A=1$) from a strongly clumped wind, for example, with half the effective scattering column ($A \approx 0.5$ at $\sim 2.5\sigma$). Testing the model's distinctive prediction (i.e., the rising amplitude with energy driven by the absorption and scattering competition) requires each band measured independently at $\gtrsim 3\sigma$ ($\sigma_A \approx 0.33$), corresponding to $\sim 7\times$ the current data for 2--4\,keV and $\sim 11\times$ for 6--8\,keV, or roughly $\sim 10\times$ overall. At this depth, the energy dependence of $A$ could be tested directly, providing a genuine test of the absorption-driven effect. Resolving the full phase structure to discriminate between wind geometries (e.g., smooth vs.\ clumped winds, focused-wind opening angle) would require tens of times the current exposure, placing this regime beyond \textit{IXPE} and into that of next-generation polarimetry missions, ideally combined with broadband coverage above 8\,keV where the predicted reflection fraction peaks.

These projections assume ideal $N^{-1/2}$ scaling and should be read as an optimistic baseline.
They assume future bins have uncertainties comparable to the current campaign and that the orbital signal remains phase-coherent across epochs (Section~\ref{sec:4.HarmReg}).
The scaling also depends on phase coverage as the predicted modulation is stronger near $\phi \approx 0.5$--$0.8$, which our current bins under-sample, so scheduling at those phases would be more efficient than adding exposure at arbitrary phases.

\section{Summary} \label{sec:7.Summary}
Reflection off the companion star and its focused stellar wind is predicted to modulate the X-ray polarization of Cygnus X-1 at half the orbital period.
We have tested this prediction using all publicly available IXPE observations (26 bins from 12 observation IDs spanning an orbital phase range of 0.030--0.988) with a simultaneous harmonic regression that models the normalized Stokes parameters $(q, u)$ as a function of both hardness ratio and orbital phase, searching at both $P_{\rm orb}/2$ and the fundamental $P_{\rm orb}$.
This model-independent approach was complemented by templates from the \texttt{SKIRT} code simulating X-ray scattering off the focused stellar wind.
Our key findings are summarized as follows:
\begin{enumerate}
    \item Neither the model-independent harmonic regression nor the model-dependent template fits show a statistically significant $P_{\rm orb}$ or $P_{\rm orb}/2$ modulation in any of the three energy bands analyzed (2--4, 4--6, and 6--8\,keV). Non-parametric permutation tests yield $p$-values above the threshold for a statistically secure detection ($p>0.01$).
    
    \item
    Inverting the injection-recovery distributions, we place 99\% confidence upper limits on the orbital modulation amplitude at $0.25\%$, $0.63\%$, and $1.96\%$ for the $P_{\rm orb}$, and $0.42\%$, $0.97\%$, and $2.78\%$ for the $P_{\rm orb}/2$ harmonic in the 2--4, 4--6, and 6--8\,keV bands, respectively.
    Both the 99\% confidence upper limits and the 99th percentiles of the phase-shuffled null distributions, which define the amplitude required to reject the null hypothesis, are given in Table~\ref{tab:upper_limits}.
    
    \item Direct template fitting yields model amplitude scaling factors of $A = 0.67 \pm 0.85$, $0.97 \pm 0.62$, and $-1.06 \pm 1.11$ at 2--4, 4--6, and 6--8\,keV, respectively. 
    These values are statistically consistent with a null result ($A=0$) within $1.6\sigma$.
    For 2--4 and 4--6\,keV, the amplitudes are consistent with the nominal stellar companion and wind-scattering model prediction ($A=1$) within $1\sigma$.
    The maximum significance is $\sim1.6\sigma$ in the 4--6\,keV band.
    This non-detection is consistent with the sensitivity limit of the current dataset, as the predicted stellar companion and wind-scattering RMS amplitudes of $\approx0.10\%$, $\approx0.33\%$, and $\approx0.49\%$ all lie below the statistical noise floor of $\sim0.18\%$, $\sim0.38\%$, and $\sim1.01\%$, respectively.
\end{enumerate}

Future observational campaigns expanding total photon statistics, ideally at the under-sampled phases $\phi \approx 0.5$--$0.8$, will enable a higher-significance test of the stellar companion and wind-scattering templates and place stricter constraints on the companion's stellar wind, offering new insights into the complex wind environments of wind-fed X-ray binaries.

\begin{acknowledgements}
The \textit{Imaging X-ray Polarimetry Explorer} (\textit{IXPE}) is a joint US and Italian mission.
The US contribution is supported by the National Aeronautics and Space Administration (NASA) and led and managed by its Marshall Space Flight Center (MSFC), with industry partner Ball Aerospace (contract NNM15AA18C).
The Italian contribution is supported by the Italian Space Agency (Agenzia Spaziale Italiana, ASI) through contract ASI-OHBI-2022-13-I.0, agreements ASI-INAF-2022-19-HH.0 and ASI-INFN-2017.13-H0, and its Space Science Data Center (SSDC) with agreements ASI-INAF-2022-14-HH.0 and ASI-INFN 2021-43-HH.0, and by the Istituto Nazionale di Astrofisica (INAF) and the Istituto Nazionale di Fisica Nucleare (INFN) in Italy.
This research used data and software products or online services provided by the IXPE Team (MSFC, SSDC, INAF, and INFN) and distributed with additional software tools by the High-Energy Astrophysics Science Archive Research Center (HEASARC), at NASA Goddard Space Flight Center (GSFC).
S.\,C., K.\,H., and H.\,K.\ acknowledge NASA for support under the grants 80NSSC24K1178, 80NSSC24K1749, and 80NSSC24K1819 and support from the McDonnell Center for the Space Sciences at Washington University in St. Louis.
B.\,V.\ acknowledges support through the European Space Agency (ESA) Research Fellowship in Space Science.
This research made use of \texttt{Astropy} \citep{astropy:2013, astropy:2018, astropy:2022}, \texttt{NumPy} \citep{NumPy2020}, \texttt{SciPy} \citep{SciPy2020}, and \texttt{Matplotlib} \citep{Matplotlib2007}.
\end{acknowledgements}

\section*{Data Availability}
All \textit{IXPE} observations used here are publicly available from the HEASARC archive.

\appendix
\section{State-Resolved Permutation Tests} \label{sec:appendixA}
To test whether the stacked null result hides a state-dependent modulation, we repeated the permutation analysis (Section~\ref{sec:4.HarmReg}) separately for the 17 hard-state and 9 soft-state bins.
The results (Table~\ref{tab:state_split}) are consistent with the null in all bands and harmonics.
The smallest $p$-value is 0.07 (hard state, 6--8\,keV, second harmonic).

\begin{table}[ht]
    \centering
    \hspace{-1.5cm}
    \begin{tabular}{lllccc}
        \hline\hline
        State & Band (keV) & Harm. & $R_{\rm obs}$ (\%) & 99\% thresh. (\%) & $p$ \\
        \hline
        \multirow{6}{*}{Soft}
          & 2--4 & 1st & 0.21 & 0.68 & 0.888 \\
          &      & 2nd & 0.42 & 0.82 & 0.428 \\
          & 4--6 & 1st & 0.66 & 1.18 & 0.311 \\
          &      & 2nd & 0.84 & 1.32 & 0.253 \\
          & 6--8 & 1st & 1.35 & 3.83 & 0.841 \\
          &      & 2nd & 2.26 & 3.96 & 0.528 \\
        \hline
        \multirow{6}{*}{Hard}
          & 2--4 & 1st & 0.20 & 0.74 & 0.927 \\
          &      & 2nd & 0.47 & 0.79 & 0.350 \\
          & 4--6 & 1st & 0.52 & 0.87 & 0.365 \\
          &      & 2nd & 0.74 & 0.95 & 0.125 \\
          & 6--8 & 1st & 1.71 & 2.50 & 0.211 \\
          &      & 2nd & 2.23 & 2.67 & 0.073 \\
        \hline
    \end{tabular}
    \caption{
    Permutation test results for the hard and soft states analyzed separately.
    No statistically significant modulation is detected in any of the three energy bands for either state.
    Columns are $R_{\rm obs}$, the 99\% detection threshold, and the permutation $p$-value, as in Table~\ref{tab:upper_limits}.
    Upper limits were not derived for the state-resolved subsets.
    All values have been recomputed with the window-averaged harmonic basis of Equation~\ref{eq:basis}.
    }
    \label{tab:state_split}
\end{table}

\section{Independence of Bins within an Observation} \label{sec:appendixB}
Our 26 bins derive from 12 observation IDs. ObsID~03002599 is treated as two separate segments (03002599a and 03002599b) because of the 63-day gap between its two epochs (Section~\ref{sec:2.Data}), giving 13 segments in total.
Bins drawn from the same observation share a calibration file and instrument configuration, so we tested whether the regression residuals are correlated within them.

We computed the standardized residuals of the simultaneous regression, $(S_i - \hat{S}_i)/\sigma_{S,i}$ for $S \in \{q,u\}$, and grouped them by segments.
Table~\ref{tab:clustering} reports the mean square of these residuals, which is unity if the measurement uncertainties fully account for the scatter, and the one-way intraclass correlation coefficient (ICC), which is zero if residuals within an observation are no more similar to each other than to residuals from other observations.

\begin{table}[ht]
    \centering
    \hspace{-1.5cm}
    \begin{tabular}{llcc}
    \hline\hline
    Harm. & Band (keV) & $\langle r^2 \rangle$ & ICC \\
    \hline
    \multirow{3}{*}{1st} & 2--4 & 1.74 & $+0.26$ \\
                         & 4--6 & 0.76 & $-0.04$ \\
                         & 6--8 & 0.78 & $+0.09$ \\
    \hline
    \multirow{3}{*}{2nd} & 2--4 & 1.68 & $+0.27$ \\
                         & 4--6 & 0.70 & $-0.04$ \\
                         & 6--8 & 0.70 & $-0.10$ \\
    \hline
    \end{tabular}
    \caption{Mean square of the standardized regression residuals ($\langle r^2 \rangle$, unity if uncertainties account for the scatter) and intraclass correlation coefficient by observation segment (ICC, zero if bins within an observation are independent).}
    \label{tab:clustering}
\end{table}

The 4--6 and 6--8\,keV bands show no evidence of clustering as the residual mean square is 0.70--0.78 and the ICC is consistent with zero.
The 2--4\,keV band behaves differently, with a residual mean square of 1.68 at $n=2$ and an ICC of $+0.27$.
We then checked whether this affects the null result.
Collapsing each observation segment to a single inverse-variance-weighted point removes within-observation correlation by construction.
Refitting the resulting 13 points against a permutation null built on the same 13 points, all bands and harmonics remain consistent with the null (Table~\ref{tab:clustering_robust}).

Note that the collapsed fit yields systematically larger $R_{\rm obs}$ than the full sample because 13 points against 8 parameters allow more noise to be absorbed into the harmonic terms.
This is a property of the reduced sample size and not evidence of signal, which is why each variant is evaluated against a permutation null built on that same reduced sample rather than against the thresholds of Table~\ref{tab:upper_limits}.
Given 13 clusters, several of which contain only one bin, these diagnostics have limited statistical power, and we present them as consistency checks rather than precise measurements.

\begin{table}[ht]
    \centering
    \hspace{-1.5cm}
    \begin{tabular}{llcccc}
        \hline\hline
        \multirow{2}{*}{Harm.} & \multirow{2}{*}{Band} &
        \multicolumn{2}{c}{Excl.\ 01002901} & \multicolumn{2}{c}{Collapsed} \\
        & (keV) & $R_{\rm obs}$ & $p$ & $R_{\rm obs}$ & $p$ \\
        \hline
        \multirow{3}{*}{1st}
          & 2--4 & 0.18 & 0.765 & 0.21 & 0.758 \\
          & 4--6 & 0.42 & 0.475 & 0.44 & 0.502 \\
          & 6--8 & 1.16 & 0.532 & 1.30 & 0.366 \\
        \hline
        \multirow{3}{*}{2nd}
          & 2--4 & 0.30 & 0.447 & 0.40 & 0.391 \\
          & 4--6 & 0.54 & 0.327 & 0.74 & 0.184 \\
          & 6--8 & 1.89 & 0.149 & 2.60 & 0.026 \\
        \hline
    \end{tabular}
    \caption{
    Robustness of the null result to within-observation correlation.
    Left: 19 bins from 12 observation segments, excluding ObsID~01002901 (only one processed with energy-scale-corrected Level-2 files).
    Right: 13 inverse-variance-weighted points, one per observation segment.
    Amplitudes in per cent.
    Each variant is evaluated against a permutation null constructed on its own sample.}
    \label{tab:clustering_robust}
\end{table}

\bibliography{sample701}{}
\bibliographystyle{aasjournalv7}

\end{document}